\documentclass[preprint]{elsarticle}

\usepackage{lineno,hyperref}

\journal{JSS, Special Issue on Program Debugging}

\usepackage{csquotes}

\usepackage[table, dvipsnames]{xcolor}
\usepackage{graphicx}
\usepackage{caption}
\usepackage{subcaption}

\usepackage{xspace}
\usepackage{amsmath}
\usepackage{amssymb}
\usepackage{booktabs}
\usepackage{numprint}
\usepackage{algorithm}
\usepackage[noend]{algpseudocode}
\usepackage[colorinlistoftodos]{todonotes} 
\usepackage{makecell}

\usepackage{color}
\usepackage{comment}

\newcommand{\change}[1]{#1}
\newcommand{\changeVtwo}[1]{#1}

\newcommand{\Break}{\State \textbf{break}}

\newcommand{\etal}{\textit{et~al.}~}
\newcommand{\tbldata}[1]{#1}

\newcommand{\type}[1]{\emph{#1}}

\newcommand{\entity}[1]{\textbf{#1}}

\newcommand{\function}[2]{\fname{#1}(#2)}
\newcommand{\fname}[1]{\textit{#1}}
\newcommand{\param}[2]{\type{#1}~\entity{#2}}

\newcommand{\spreadsheetfunction}[1]{\textsc{#1}}

\newcommand{\fritz}{\textsc{Fritz}\xspace}

\newcommand{\rqone}{Can existing smell detection techniques be improved by applying structural information in the smell's detection process?}
\newcommand{\rqtwo}{Are novel spreadsheet smell detection techniques that are based on structural information able to detect new quality issues, and do they perform similar to traditional ones?}

\newtheorem{definition}{Definition}
\newtheorem{example}{Example}

\npdecimalsign{.}

\begin{document}

\begin{frontmatter}

\title{On the Refinement of Spreadsheet Smells by means of Structure Information}

%
%

\author{Patrick Koch\fnref{adress}}
\ead{Patrick.Koch@aau.at}
\author{Birgit Hofer\fnref{adressTU}}
\ead{bhofer@ist.tugraz.at}
\author{Franz Wotawa\fnref{adressTU}}
\ead{wotawa@ist.tugraz.at}

\fntext[adress]{Klagenfurt University, Austria}
\fntext[adressTU]{Graz University of Technology, Austria}

\begin{abstract}
\change{
Spreadsheet users are often unaware of the risks imposed by poorly designed spreadsheets.
One way to assess spreadsheet quality is to detect smells which \changeVtwo{attempt to} identify parts of spreadsheets that are hard to comprehend or maintain and which are more likely to be the root source of bugs.
Unfortunately, current spreadsheet smell detection techniques suffer from a number of drawbacks that lead to incorrect or redundant smell reports.
For example, the same quality issue is often reported for every copy of a cell, which may overwhelm users.
To deal with these issues, we propose to refine spreadsheet smells by exploiting inferred structural information for smell detection.
We therefore first provide a detailed description of our static analysis approach to infer clusters and blocks of related cells.
We then elaborate on how to improve existing smells by providing three example refinements of existing smells that incorporate information about cell groups and computation blocks.
Furthermore, we propose three novel smell detection techniques that make use of the inferred spreadsheet structures.
Empirical evaluation of the proposed techniques suggests that the refinements successfully reduce the number of incorrectly and redundantly reported smells, and novel deficits are revealed by the newly introduced smells.
}
\end{abstract}


\end{frontmatter}


\section{Introduction}
End-user programming has gained a lot of attention. 
Without doubt, spreadsheets are the most prominent example of end-user programs.
They are intuitive to use and widespread. Nearly everybody has access to them via spreadsheet environments like Microsoft Excel, Google Sheets, and Numbers.
Professionals, managers, sales \change{people}, and administrators use spreadsheets exhaustively~\cite{ScaffidiSM05}, and often base important decisions on them.
Their easy usage is one of the main reasons why spreadsheets are so popular.

On the downside, there is a lack of quality awareness.
The ease of use often prevents people from taking training courses and lets them handle spreadsheets in a learning by doing style.
Therefore, many people are not aware of the risks that come with spreadsheets.
The list of horror stories\footnote{see \url{http://www.eusprig.org/horror-stories.htm}} where spreadsheet errors caused significant reputation loss or immense financial damages is long, as the following three examples illustrate:
(1)~JP Morgan lost \$\,400 million because of a fault in their value-at-risk model spreadsheet~\cite{taskForce}. 
(2)~The US-village West Baraboo has to pay more than \$\,400,000 in additional interests because of a spreadsheet error~\cite{westbarboo}.
(3)~The Canadian power generation company TransAlta lost \$\,24 million
because of a copy-paste error in a spreadsheet that led them to buy more US power transmission hedging contracts than necessary~\cite{transalta}.
One might think that these examples are just extraordinary exceptions, but field studies \cite{corr/Panko16,Powell2009} confirm a consistent lack of quality in business spreadsheets.

To address such quality issues, we are working among a large number of fellow researchers to develop quality assurance (QA) techniques for spreadsheets (see Jannach \textit{et al.}~\cite{JannachSHW14} for an overview).
As spreadsheets \changeVtwo{can be regarded as} a form of software, as pointed out by Hermans \textit{et~al.}~\cite{HermansJRASH16}, a number of QA techniques for general software development were adopted for the domain.
One such technique is the detection of spreadsheet smells that were adopted from code smells~\cite{Fowler99}.
A code smell indicates lacks in the quality of a part of source code.
This part could either be hard to comprehend, hard to maintain, or error-prone.
\changeVtwo{
Spreadsheet smells similarly attempt to infer deficient parts in spreadsheets~\cite{JansenH15}.}
There are three basic types of spreadsheet smells:
(1) input smells~\cite{CunhaFRS12} indicate, for example, missing input values, wrong numeric values, and mistyped strings;
(2) formula smells~\cite{HermansPD12a} relate to complex formulas, long calculation chains, and duplicated formulas; and
(3) inter-worksheet smells~\cite{HermansPD12} point out cells and worksheets that excessively refer to other worksheets, or which are excessively referred to by other worksheets.


In this work, we regard spreadsheet smells to be used as part of a checking regime for spreadsheets.	
In a survey, Mireille Ducass{\'{e}}~\cite{duc93} classified debugging approaches where one method is checking. 
In checking, patterns in programs are used for identifying potential bug locations. 
Such patterns are code smells, which directly point to suspicious parts of the program and therefore potentially locate bugs. 
Debugging can be seen as an activity comprising fault detection, fault localization, and fault correction to eliminate bugs in software.
In checking, fault detection and fault localization are preformed at the same time using patterns that are classifying a code smell.
\changeVtwo{Similarly, we regard the use of spreadsheet smells to point out suspicious parts of a spreadsheet that a user should examine preferrentially when checking for correct functionality of said sheet.}

\changeVtwo{
We would however like to point out, that we currently lack substantive empirical evidence that clearly links spreadsheet smells to fault proneness, and we thus regard this as an opportunity for future work.
Nevertheless, we still suspect our perception to be valid, as a recent case study linked a reduction in spreadsheet smell detections with a subjective impression of heightened quality and maintainability of spreadsheets \cite{JansenH15}, and a number of previous works successfully used}
spreadsheet smells in the context of spreadsheet debugging, including the FaultySheet Detective tool introduced by Abreu \textit{et~al.}~\cite{AbreuCFMPS14a}, as well as Table Clones and other approaches discussed by Dou \textit{et~al.}~\cite{DouCW14, Dou2016}. 

\change{
Unfortunately, spreadsheet smell detection suffers from certain weaknesses and blind spots.
For example, duplication is a common problem: 
if a smelly cell of a spreadsheet is copied, its copies are often also reported as being smelly, and seeing the same smell being reported over and over again can be very frustrating for a user.
Other weaknesses affect only certain smells.
For example, the input smell \enquote*{Pattern Finder}~\cite{CunhaFMMS12,CunhaFRS12} detects deviating cell types for cells in the same row or column.
However, the reported smelly cells often turn out to be false positives, because not all cells in a row or column serve the same purpose.
}

\change{
We propose to compensate existing shortcomings by incorporating structural information in the detection procedures of spreadsheet smells.
As shown by a study of Cunha \textit{et~al.}~\cite{CunhaFMS15}, providing abstract structure information can enhance the ability of end-users to understand and efficiently use spreadsheets.
The detection of spreadsheet structures follows the approach we outlined in previous work~\cite{iwpd/KochHW16}, where we explained and evaluated the static analysis process.
As the targeted smell refinements require detailed information about the inferred structures, the current work describes the structure analysis in greater detail, including a description of the handling of cell references and area references, as well as the theory of absolute versus relative references.
}

\change{
Inferred structure information (input groups, formula groups, computation blocks, and headers of such blocks) can be applied in several ways for smell refinement, as we demonstrate in three examples:
(1) for the \emph{Pattern Finder} input smell, we check for deviations only within groups instead of checking complete rows and columns;
(2) for the \emph{Long Calculation Chain} formula smell, we report a smell only once per group instead of reporting it for each individual cell; and
(3) for \emph{Feature Envy} inter-worksheet smell, we count the connections between worksheets for each group instead of for each cell.
Moreover, structural information can be used to check for novel quality issues.
We demonstrate this by introducing three new spreadsheet smell detection techniques which are based on the results of the structural analysis: 
(1) \emph{Overburdened Worksheet} indicates worksheets that contain too much functionality;
(2) \emph{Inconsistent Formula Group Reference} signals inconsistencies within the references of groups of formulas; and
(3) \emph{Missing Header} indicates gaps in series of header cells.
}

\change{
We investigated the performance of the improved and new smells using an empirical evaluation on a well known dataset in combination with a manual investigation of detected smells on a selected subset of spreadsheets.
This investigation revealed that the improved smell versions are successful in limiting false positive and duplicate detections, and the new smells can be used in combination with existing techniques to point out novel quality issues.
}

The remainder of this paper is organized as follows:
Section~\ref{sec:ex} pictures a motivating example for the application of the proposed refinements, and 
Section~\ref{sec:prelim} defines the most important concepts of spreadsheets used in this paper.
Section~\ref{sec:structure} explains how our structural analysis approach works.
Section~\ref{sec:smellImprovements} shows how three existing smell detection techniques can be improved by means of structure information and introduces three novel smell detection techniques that make use of structures.
Section~\ref{sec:eval} evaluates the improved and new techniques, 
Section~\ref{sec:rw} discusses related work, 
and Section~\ref{sec:conclusion} concludes this paper.

\section{Motivating Example}\label{sec:ex}
\begin{figure}[htp]
	\centering
	\begin{subfigure}[b]{1.0\textwidth}
		\includegraphics[width=\textwidth]{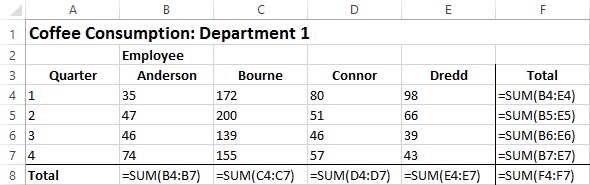}
		\caption{Department1}
		\label{fig:example_coffee_worksheet_department1_formulas}
	\end{subfigure}
	   
	\begin{subfigure}[b]{1.0\textwidth}
		\includegraphics[width=\textwidth]{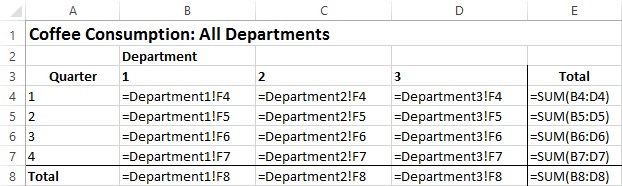}
		\caption{Total}
		\label{fig:example_coffee_worksheet_total_formulas}
	\end{subfigure}
	    
	\begin{subfigure}[b]{1.0\textwidth}
		\includegraphics[width=\textwidth]{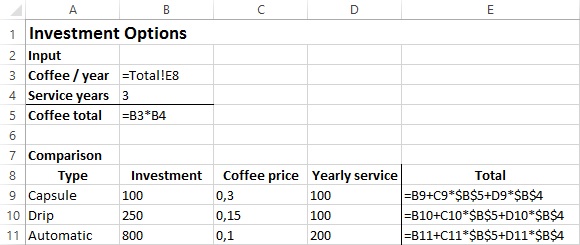}
		\caption{Investment}
		\label{fig:example_coffee_worksheet_investment_formulas}
	\end{subfigure}
	\caption{Formula views of the different worksheets of our running example}
	\label{fig:example_coffee_formulas}
\end{figure}

The spreadsheet illustrated in Figure~\ref{fig:example_coffee_formulas} compares purchase options for a new office coffee machine.
The coffee machine should be chosen from one of three alternatives (capsule, drip, automatic) in such a way that the total costs over a period of three years are the lowest.
For this, the coffee consumption per department is captured in worksheets \texttt{Department1/2/3} (see Figure~\ref{fig:example_coffee_worksheet_department1_formulas}, the non-illustrated worksheets are similar), summed up in worksheet \texttt{Total} (see Figure~\ref{fig:example_coffee_worksheet_total_formulas}), and the resulting cost of the three alternatives compared in worksheet \texttt{Investment} (see  Figure~\ref{fig:example_coffee_worksheet_investment_formulas}).

Assume a smell detection technique reports cell E9 of the  \texttt{Investment} worksheet as smelly, because E9's formula is too complex.
Since the cells E10 and E11 have the same complex formula, they would also be reported as smelly.
When only these three smells are reported, then a user would probably immediately see that these cells suffer from the same smell.
\change{
However, when faced with an abundance of smell reports in a larger spreadsheet, the connection between related smelly cells is harder to comprehend for a user.
In contrast, our proposed smell refinement detects and indicates related smelly cells as one smelly unit instead, providing concise feedback to the user.
}

\section{Preliminaries}\label{sec:prelim}

A spreadsheet consists of a set of worksheets; a worksheet consists of set of cells.
A cell \entity{c} has a content and can be uniquely identified by its coordinates and the worksheet it belongs to:
The function \function{x}{\param{cell}{c}} returns the column index of cell~\entity{c} within a worksheet; \function{y}{\param{cell}{c}} returns its row index~\cite{HoferRWAG13}.
%
%
%
%
We can use the cells' coordinates to determine neighboring cells.

\begin{definition}[Neighbors]\label{def:neighbors}
Two cells, \entity{c} and \entity{c'}, are neighbors if their Manhattan distance \change{(simple sum of the differences of the respective column and row indices)} is one.
The function \function{neighbors}{\param{cell}{c}} returns c's neighbors~\cite{iwpd/KochHW16}:
$\function{neighbors}{c}=\{c' \in \function{cells}{\function{ws}{c}} : |x(c')-x(c)|+|y(c')-y(c)|=1\}.$
\end{definition}

\begin{example}\label{ex:neighbors}
Cell B5 of the worksheet from Figure~\ref{fig:example_coffee_worksheet_investment_formulas} has as neighbors the cells B4, B6, A5, and C5.
\end{example}

\begin{definition}[Connected]\label{def:connected}
The function \function{connected}{\param{\{cell\}}{C}} checks whether a set of cells \entity{C} is completely connected by neighbourhood relations; it returns true, if $\{|C|=1 \vee (\exists c \in C: \exists c' \in C \text{ with } c' \in \textit{neighbors}(c) \wedge \function{connected}{C\setminus c})\}$. 
\end{definition}
All cells in a set of connected cells must reach every other cell in the set by following only neighborhood relations of cells within the set.

Beside its coordinates, each cell has a type depending on its content.
This can be either formula, Boolean, numeric, string, error, or empty.
\enquote*{Formula} is a dominant type: formula cells compute to other cell types, however, the cell's type is still \enquote*{formula}.
The function \function{type}{\param{cell}{c}} returns the type of a cell.

\begin{example}\label{ex:formulaType}
Although cell F4 of worksheet \texttt{Department1}  (see Figure~\ref{fig:example_coffee_worksheet_department1_formulas}) computes a numerical value, its \textit{type} is \enquote*{formula}.
\end{example}

Formula cells might reference other cells.
References can either be relative or absolute:
a relative reference determines the referenced cell in relation to the position of the referencing cell;
an absolute reference determines the referenced cell by indicating its position within the worksheet independent of the position of the referencing cell.
There are two notations for referencing cells, the A1 and the R1C1 notation.
Both support relative and absolute references.

In the A1 notation, a cell \entity{c} references another cell \entity{c'} by indicating the absolute position of \entity{c'}, i.e., \function{x}{\entity{c'}} and \function{y}{\entity{c'}}. Absolute references place a \$ sign preceding the coordinate position. 
%
Absolute references in A1 notation are relevant when copying cells.
For example, if the relative cell reference \enquote*{A1} is copied from cell \texttt{B1} to cell \texttt{B2}, the reference will change, according to the shift of the cell position, to \enquote*{A2}. In contrast, an absolute cell reference \enquote*{\$A\$1}, when copied from any cell to any other cell, will always refer to the coordinates A1. 
Cell references can be absolute (e.g., \enquote*{\texttt{\$A\$1}}), relative (e.g., \enquote*{\texttt{A1}}), or mixed (e.g., \enquote*{\texttt{\$A1}}, \enquote*{\texttt{A\$1}}).
Names of worksheets are only included when the worksheet of the referenced cell differs from that of the referencing cell.

\begin{definition}[R1C1 notation]\label{def:r1c1Notation}
In the R1C1 notation, a cell references another cell by indicating the relative position of the referenced cell. The relative position of cell $c'$ with respect to $c$ is indicated as \texttt{R[}\function{y}{c'} - \function{y}{c}\texttt{]C[}\function{x}{c'} - \function{x}{c}\texttt{]}.
Absolute references to a cell $c'$ are written as \texttt{R}$\function{y}{c'}$\texttt{C}$\function{x}{c'}$.
The names of worksheets are only indicated when the worksheet of the referenced cell differs from that of the referencing cell.~\cite{iwpd/KochHW16}
\end{definition}
 
If a cell refers to a cell in the same row or column, the \enquote*{0} for indicating the row or column index can be left out, e.g. \enquote*{\texttt{RC[-1]}} for indicating the cell above and \enquote*{\texttt{R[-1]C}} for indicating the left cell.
References to absolute row or column positions omit the square brackets, e.g. \enquote*{\texttt{R1C1}} indicates cell \texttt{A1}, the first cell in the first column.

\begin{definition}[Formula]\label{def:formula}
The function \function{formula}{\param{cell}{c}} returns the formula of cell~\entity{c} in R1C1 notation if \entity{c} is a formula cell;
otherwise it returns \texttt{void}.
\end{definition}

\begin{example}\label{ex:r1c1}
Cell B5's formula of the worksheet \texttt{Investment} (Figure~\ref{fig:example_coffee_worksheet_investment_formulas}) is \enquote*{\texttt{=B3*B4}} in A1 notation and \enquote*{\texttt{=R[-2]C*R[-1]C}} in R1C1 notation.
Cell B3's formula of the same worksheet is \enquote*{\texttt{=Total!E8}} in A1 notation and \\ \enquote*{\texttt{=Total!R[5]C[3]}} in R1C1 notation.
\end{example}

Frequently, formulas are copied to other cells. 
Many spreadsheet environments support copying by drag\&drop.
Even though the copied cells reference other cells than the original cells, they are semantically equal.
We identify semantically equal cells by means of copy-equivalence:

\begin{definition}[Copy-equivalence]\label{def:cop-eq}
Two cells \entity{c}, \entity{c'} are copy-equivalent if their formulas in R1C1 notation are identical ($\function{formula}{c}\equiv\function{formula}{c'})$.
\end{definition}

\begin{example}\label{ex:copyEquiv}
For the spreadsheet from Figure~\ref{fig:example_coffee_formulas}, we identify the cells E9, E10, and E11 of the worksheet \texttt{Investment} as copy-equivalent since they have the same formula (\enquote*{\texttt{=RC[-3]+RC[-2]*R5C2+RC[-1]*R4C2}}) in R1C1 notation. The other copy-equivalent cells are
\begin{itemize}
	\item B8, C8, D8, E8, and F8 of the worksheets \texttt{Department1}, \texttt{Department2} and \texttt{Department3}  (\enquote*{\texttt{=SUM(R[-4]C:R[-1]C)}}),
	\item F4, F5, F6, and F7  of \texttt{Department1}, \texttt{Department2} and \texttt{Department3} (\texttt{=SUM(RC[-4]:RC[-1])})
	\item B4, B5, B6, B7, and B8 of worksheet \texttt{Total} (\enquote*{\texttt{=Department1 RC[4]}})
	\item C4, C5, C6, C7, and C8 of worksheet \texttt{Total} (\enquote*{\texttt{=Department2 RC[3]}})
	\item D4, D5, D6, D7, and D8 of worksheet \texttt{Total} (\enquote*{\texttt{=Department3 RC[2]}})
	\item E4, E5, E6, E7, and E8 of worksheet \texttt{Total} (\enquote*{\texttt{=SUM(RC[-3]:RC[-1])}})
\end{itemize}
B3 and B5 of  worksheet \texttt{Investment} do not have any copy-equivalent cells.
\end{example}

Since references can be absolute, relative or mixed, we separately consider the x and y coordinates when determining the position of a referenced cell:

\begin{definition}[Coordinate reference]\label{def:coordinateReference}
A coordinate reference \entity{r$_{c}$} represents either the x or the y coordinate of cell $c$ and consists of a value and a type.
The function \function{absolute}{\param{coordinate reference}{r$_{c}$}} returns \texttt{true} if \entity{r$_{c}$} is an absolute reference; it returns \texttt{false} if \entity{r$_{c}$} is a relative reference.
The function \function{value}{\param{coordinate reference}{r$_{c}$}} returns 
for an absolute reference, the absolute index of an absolute reference;
for a relative reference, it returns the deviation of the reference from a base index.
\end{definition}

\begin{example}\label{ex:coordinate reference}
The reference \enquote*{\texttt{R4C[2]}} features two coordinate references. Its row reference, \enquote*{\texttt{R4}}, is absolute and refers to the fourth row of the worksheet. Its column reference, \enquote*{\texttt{C[2]}}, is relative and refers to the column two positions right of the column of the cell containing the reference.
\end{example}

\begin{definition}[Dereferencing coordinates]\label{def:deref_coordinateReference}
The function \function{derefCoordinate}{\param{ coordinate reference}{r$_{c}$}, \param{index}{coordinate}} returns the dereferenced index of a coordinate reference \entity{r$_{c}$} in the formula of the cell at index \entity{coordinate}:

\function{derefCoordinate}{r$_{c}$, coordinate} = 
$\begin{cases}
\function{value}{r_{c}} & \text{if} ~ \function{absolute}{r_{c}}\\
coordinate + \function{value}{r_{c}} & \text{otherwise}.\\
\end{cases}$
\end{definition}

\begin{example}\label{ex:coordinateReference}
Cell E9's formula of the worksheet \texttt{Investment} (Figure~\ref{fig:example_coffee_worksheet_investment_formulas}) is \enquote*{\texttt{=B9+C9*\$B\$5+D9*\$B\$4}} in A1 notation and \enquote*{\texttt{=RC[-3]+RC[-2]*R5C2+RC[-1]* R4C2}} in R1C1 notation. 
Its first reference, \enquote*{\texttt{RC[-3]}} (\enquote*{\texttt{B9}} in A1 notation), is a combination of a relative column-reference with value -3 and a relative row-reference with value 0.
Its third reference, \enquote*{\texttt{R5C2}} (\enquote*{\texttt{\$B\$5}} in A1 notation), is a combination of an absolute column-reference and an absolute row-reference.
\end{example}

Formulas may contain two different types of references: cell references and area references.
Cell references, hereinafter referred to as \enquote{references}, point to individual cells;
area references point to sets of cells.
Cell references consist of two parts:
(1)~the target worksheet of the reference (provided by the function \function{ws}{\param{reference}{r}}) and
(2)~the coordinate references in column and row orientation (provided by the functions \function{x}{\param{reference}{r}} and \function{y}{\param{reference}{r}}).

\begin{definition}\label{deref_cellReference}
The function \function{deref}{\param{reference}{r}, \param{cell}{c}} resolves a reference~\entity{r} contained in the formula of cell~\entity{c}, returning the referred cell~\entity{c'}:
\begin{equation}
\begin{aligned}
\nonumber
\function{deref}{r, c} = c' \in \function{cells}{\function{ws}{r}} :~&  \function{x}{c'} \equiv \function{derefCoordinate}{\function{x}{r}, \function{x}{c}}~ \wedge \\
&\function{y}{c'} \equiv \function{derefCoordinate}{\function{y}{r}, \function{y}{c}}
\end{aligned}
\end{equation}
The function \function{refs}{\param{formula}{f}} returns the set of all cell references of formula \entity{f}. 
\end{definition}


In contrast to cell references, area references refer to a set of cells in a rectangular area of a worksheet.
The rectangular area is described by its top-left start coordinates and its bottom-right end coordinates.
Area references consist of three parts:
(1)~the worksheet to which the reference \entity{r$_{a}$} refers to (accessibly by the function \function{ws}{\param{area reference}{r$_{a}$}}),
(2)~the x and y coordinate references of the start coordinates of the area reference (accessible by the functions \function{x$_{1}$}{\param{area reference}{r$_{a}$}} and \function{y$_{1}$}{\param{area reference}{r$_{a}$}}), and
(3)~the x and y coordinate references of the end coordinates of the area reference (accessible by the functions \function{x$_{2}$}{\param{area reference}{r$_{a}$}} and \function{y$_{2}$}{\param{area reference}{r$_{a}$}}).

\begin{definition}\label{deref_areaReference}
Function \function{deref}{\param{area reference}{r$_{a}$}, \param{cell}{c}} resolves an area reference \entity{r$_{a}$} contained in the formula of cell \entity{c}, returning the set of referred cells:
\begin{equation}
\nonumber
\function{deref}{r_{a}, c} =
\left\{  
	c' \in \function{cells}{\function{ws}{r_{a}}} : \left| 
		\begin{array}{l}
			\function{derefCoordinate}{\function{x$_{1}$}{r_{a}}, \function{x}{c}} \leq \function{x}{c'}~\wedge \\
			\function{derefCoordinate}{\function{x$_{2}$}{r_{a}}, \function{x}{c}} \geq \function{x}{c'}~\wedge \\
			\function{derefCoordinate}{\function{y$_{1}$}{r_{a}}, \function{y}{c}} \leq \function{y}{c'}~\wedge \\
			\function{derefCoordinate}{\function{y$_{2}$}{r_{a}}, \function{y}{c}} \geq \function{y}{c'} \\
		\end{array} 
	\right. 
\right\}
\end{equation}
Function \function{refs$_{a}$}{\param{formula}{f}} returns the set of all area references of formula \entity{f}. 
\end{definition}


\begin{definition}[Referenced cells]\label{def:referencedCells}
The function \function{$\rho$}{\param{cell}{c}} returns the set of cells that are referenced in \entity{c}'s formula \entity{f}:
$$\function{$\rho$}{c} = \ \bigcup_{r_{c} \in \function{refs}{f}}\{\function{deref}{r_{c}, c}\} \cup \bigcup_{r_{a} \in  \function{refs$_{a}$}{f}}\function{deref}{r_{a}, c}.$$
\end{definition}

\begin{example}\label{ex:refCells}
For the cell E11 in the worksheet \texttt{Investment}, we compute the following reference information:
\function{refs}{\texttt{Investment!E11}} = \{\texttt{RC[-3]}, \texttt{RC[-2]}, \texttt{R5C2}, \texttt{RC[-1]}, \texttt{R4C2}\}, 
\function{refs$_{a}$}{\texttt{Investment!E11}} = \{\}, and 
\function{$\rho$}{\texttt{Investment!E11}} = \{\texttt{B11}, \texttt{C11}, \texttt{B5}, \texttt{D11}, \texttt{D4}\}.
For the cell \texttt{Total!E8}, we compute the following reference information:
\function{refs}{\texttt{Total!E8}} = \{\}, \function{refs$_{a}$}{\texttt{Total!E8}} = \{\texttt{RC[-3]:RC[-1]}\}, and 
\function{$\rho$}{\texttt{Total!E8}}=\{\texttt{B8}, \texttt{C8}, \texttt{D8}\}.
\end{example}

\section{Structure Analysis}\label{sec:structure}
\change{
This section describes our approach to infer structure information from a spreadsheet, that is then used to improve and create spreadsheet smells.
The general approach was already presented in our IWPD paper~\cite{iwpd/KochHW16}.
In the present work, we describe the analysis process in greater detail. 
We provide examples for each analysis step by referring to the running example in Section~\ref{sec:ex}, and also elaborate on previously unattended topics that are relevant for the definition of smell refinements and novel, structure-based smells (e.g. handling of cell and area references).
}

The goal of the structural analysis is to identify input groups, formula groups, computation blocks, and headers.
Groups are one-dimensional areas in a worksheet.
Cells belonging to the same group share the same purpose, e.g. summing up data, even though they might operate on different input data.
Cells of an input group provide their value to formula groups, but do not themselves refer to any other group.
Cells of formula groups process the same calculation on different input data and refer to input groups or intermediate formula groups.  
Computation blocks are rectangular areas containing related input groups, formula groups, and empty cells.
Headers serve as labels for rows and columns of computation blocks.

\begin{example}\label{ex:whatIsDected}
For the worksheet \texttt{Department1} from Figure~\ref{fig:example_coffee_worksheet_department1_formulas}, the structural analysis process will detect input groups for the area \texttt{B4:E7}, two formula groups (\texttt{B8:F8} and \texttt{F4:F7}), and one computation block (\texttt{B4:F8}).
The cells in the area \texttt{A4:A8} are the row headers for this block and the cells \texttt{B3:F3} are the column headers.
\texttt{A3} and \texttt{B2} are meta headers.
While all cells in the area \texttt{A4:E7} are numerical values, the cells \texttt{A4:A7} have a different purpose then the cells \texttt{B4:E7}. 
\end{example}

\begin{figure}[ht]
	\centering
	\includegraphics[width=0.83\textwidth]{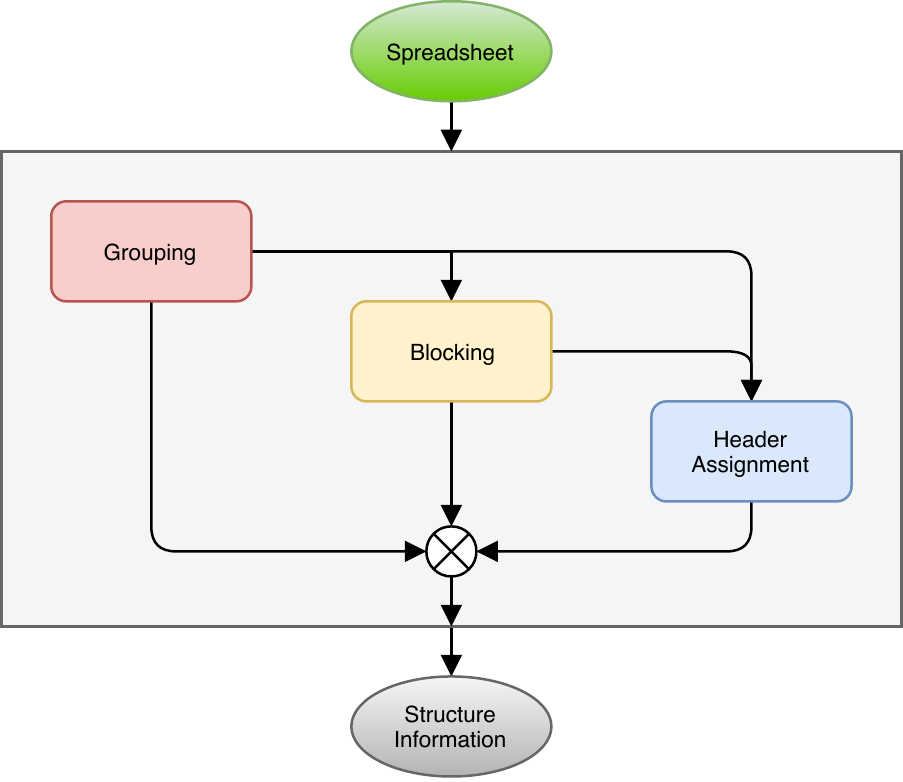}
	\caption{Block creation. Gray cells represent grouped cells, red cells represent non-blockable cells (i.e., non-empty cells that are not part of any group), and frames represent blocks.}\label{fig:analysisProcess}
\end{figure}

\change{
Figure~\ref{fig:analysisProcess} illustrates the overall analysis process:
(1)~\emph{Grouping} finds group of related cells, based on various criteria;
(2)~\emph{Blocking} combines groups into cohesive blocks; and
(3)~\emph{Header Assignment} relates header cells to blocks and the contained groups.
Each step in the process infers specific structural properties of the input sheet, and includes information from preceding process steps for in the analysis.
}

\subsection{Grouping}
\change{
In the grouping step, we infer different groups the cells in each worksheet, according to their type (string, numeric, formula, \dots), content (in case of formula cells), and position within the worksheet.
The first action within the Grouping step is to determine \emph{type-based groups} within each worksheet.
}

\begin{definition}[Type-based group]\label{def:typeBasedGrouping}
A type-based group \entity{C} consists of a set of cells which
(1)~have the same cell type, 
(2)~are connected (see Def.~\ref{def:connected}), and 
(3)~feature the same formula in R1C1 notation (if the cells are of type formula):
\[
	\\\function{typeBasedGroup}{C} = 
	\begin{cases}
		\text{true} & \text{if } (
		\{ \forall{c,c' \in C}: \function{type}{c}=\function{type}{c'}\\
		& \wedge \function{formula}{c}\equiv\function{formula}{c'}\}\\
		& \wedge \function{connected}{C} )\\
		\text{false} & \text{otherwise}.
	\end{cases}
\]
\end{definition}

\begin{example}\label{ex:typeBasedGrouping}
Worksheet \texttt{Department1} (Figure~\ref{fig:example_coffee_formulas}) contains the following type-based groups:
\{\texttt{A1}\}, \{\texttt{B2;A3:F3}\}, \{\texttt{A4:E7}\}, \{\texttt{F4:F7}\}, \{\texttt{A8}\}, and \{\texttt{B8:F8}\}.
Worksheet \texttt{Total} has \{\{\texttt{A1}\}, \{\texttt{B2}\}, \{\texttt{A3}\}, \{\texttt{B3:D3}\}, \{\texttt{E3}\}, \{\texttt{A4:A7}\}, \{\texttt{A8}\}, \{\texttt{B4:B8}\}, \{\texttt{C4:C8}\}, \{\texttt{D4:D8}\}, \{\texttt{E4:E8}\}\} as type-based groups.
Worksheet \texttt{Investment} has \{\{\texttt{A1:A5}\}, \{\texttt{B3}\}, \{\texttt{B4}\}, \{\texttt{B5}\}, \{\texttt{A7:A11;B8:E8}\}, \{\texttt{B9:D11}\}, \{\texttt{E9:E11}\}\} as type-based groups.
\end{example}

In the example, the headers in the cells \texttt{A4:A7} of the worksheet \texttt{Department1} are in the same group as the input cells \texttt{B4:E7}.
This renders header inference problematic.
\change{
The header and input cells will be separated in the subsequent analysis, by examining which cells are referenced by formula cells.
}

The type-based groups mainly serve as starting point for further processing of related formula cells.
In the next step, we synthesize the formula cells of each worksheet into cohesive units.
For doing so, we define \emph{formula groups}.

\begin{definition}[Formula group]\label{def:FormulaGroup}
A type-based group \entity{C} is a formula group if the group consists of formula cells:
$$\function{formulaGroup}{C}= 
\begin{cases}
		\text{true}  & \text{if } \function{typeBasedGroup}{C}=\text{true} \\
		             & \wedge (\forall c \in C: \function{type}{c}=\text{\enquote*{formula}})\\
		\text{false} & \text{otherwise}.
\end{cases}
$$
The function \function{formulaGroups}{\param{worksheet}{w}} returns for a worksheet \entity{w} the set of all formula groups contained in \entity{w}.
The function \function{formula}{\param{formula group}{C}} returns the formula of group \entity{C}.
\end{definition}
\begin{example}\label{ex:FormulaGroup}
The function \function{formulaGroups}{\texttt{Department1}} returns \{\{\texttt{B8:F8}\}, \{\texttt{F4:F7}\}\},
\function{formulaGroups}{\texttt{Total}} returns \{\{\texttt{B4:B8}\}, \{\texttt{C4:C8}\}, \{\texttt{D4:D8}\}, \{\texttt{E4:E8}\}\}, and 
\function{formulaGroups}{\texttt{Investment}} returns \{\{\texttt{B3}\}, \{\texttt{B5}\}, \{\texttt{E9:E11}\}\}.
\end{example}

\change{
Where necessary, formula groups are then divided into one-dimensional partitions: 
\emph{partitioned formula groups} are one-dimensional areas, i.e., they either have a row-wise or column-wise orientation.
}
When partitioning a formula group, we choose the orientation of the partitions in such a way that the smallest number of partitioned formula groups is created.
In case of a draw, we opt for column-oriented groups because our analysis of a public spreadsheet corpus \change{(EUSES, see Section~\ref{sec:eval} for further information)} concluded that 34\,\% of worksheets contain references to column-oriented areas, whereas only 13\,\% feature references to row-oriented areas. Thus, it is more likely that a column-oriented group was applied by the spreadsheet's author.

\begin{definition}[Partitioned formula group]\label{def:PartitionedFormulaGroup}
A formula group \entity{C} is a partitioned formula group \entity{g}, if its cells form a one-dimensional group:
\[
	\textit{PFGroup}(C) = 
	\begin{cases}
		\text{true}  & \text{if}~\function{formulaGroup}{C} \wedge \{\{\forall{c,c' \in C}: \function{x}{c}=\function{x}{c'}\} \vee \\ 
					 & \{\forall{c,c' \in C}: \function{y}{c}=\function{y}{c'}\}\}\\
		\text{false} & \text{otherwise}.
	\end{cases}
\]
The function \function{partFormulaGroups}{\param{worksheet}{w}} returns all partitioned formula groups of worksheet \entity{w}.
\end{definition}

\begin{example}\label{ex:PartitionedFormulaGroup} 
Since all formula groups in Example~\ref{ex:FormulaGroup} are one-dimensional, 
the partitioned formula groups are the same. 
\end{example}

\change{
After processing all partitioned formula groups of a spreadsheet, we first determine which \emph{referred formula groups} are connected to each partitioned formula group via references from the cells within a group.
}


\begin{definition}[Referred formula groups]\label{def:FormulaGroupConnection}	
	A partitioned formula group \entity{g} has a group reference to another partitioned formula group \entity{g'}, if any of its cells refer to any cell within \entity{g'}:	
	\[
	\textit{refersTo}(g,g') = 
	\begin{cases}
	\text{true}  & \text{if}~\exists c: c \in g \wedge \function{$\rho$}{c} \cap g' \neq \emptyset \\
	\text{false} & \text{otherwise}.
	\end{cases}
	\]
	The function \function{referredFormulaGroups}{\param{Partitioned formula group}{g}} returns the set of all partitioned formula groups to which \entity{g} refers to.	
\end{definition}

\begin{example}\label{ex:FormulaGroupReference}
	The \texttt{Department1} worksheet has two partitioned formula groups: \texttt{B8:F8} and \texttt{F4:F7}. 
	Cell F8 in group \texttt{B8:F8} refers to cells in group \texttt{F4:F7}; the function \function{refersTo}{\texttt{B8:F8},\texttt{F4:F7}} returns true.	
	Hence, the function \function{referredFormulaGroups}{\texttt{B8:F8}} returns \{\texttt{F4:F7}\}. 
	Group \texttt{F4:F7} does not refer to cells of any other partitioned formula group (\function{referredFormulaGroups}{\texttt{F4:F7}}=$\emptyset$).
\end{example}

\change{
Next, we establish \emph{reference-based groups} using the references of each formula group. We thus identify connected areas of cells that serve as input for a specific calculation in the spreadsheet (a partitioned formula group).
}

\begin{definition}[Reference-based group]\label{def:refBasedGroup}
A reference-based group \entity{g$_{r}$} is a set of cells that are referred to by a partitioned formula group \entity{g}. 
Each group \entity{g$_{r}$} can be attributed to either a specific reference \entity{r} or a specific area reference \entity{r$_{a}$} of~\entity{g}.
In case of a reference, \entity{g$_{r}$} is the collection of all cells that are referred to by any cell of \entity{g} via the reference \entity{r}.
In case of an area reference, \entity{g$_{r}$} is a collection of all cells that are referred to by a specific cell of \entity{g} via the area reference \entity{r$_{a}$}.
The function \function{referenceGroups}{\param{partitioned formula group}{g}} returns the set of reference-based groups \{\entity{g$_{r}$}\} for a partitioned formula group \entity{g}.
\[
	\textit{referenceGroups}(g) = \function{refGroups}{g} \cup \function{areaRefGroups}{g}
\]
\[
	\textit{refGroups}(g) = \bigcup_{r \in \function{refs}{f}} \left\{ 
		\bigcup_{c \in {g}} \{\function{deref}{r, c}\}    
	\right\}
\]
\[
	\textit{areaRefGroups}(g) = 
	\bigcup_{r_a \in \function{refs$_a$}{f}}
	\left\{\bigcup_{c\in g}\function{deref}{r_{a}, c}\right\}
\]
where $f=\function{formula}{g}$.
The function \function{referenceGroups}{\param{worksheet}{w}} returns for a worksheet \entity{w} the set of all reference-based groups of \entity{w}.
\end{definition}

\begin{example}\label{ex:refBasedGroup}
The formula of the partitioned formula group \texttt{Investment!\\E9:E11} (Figure~\ref{fig:example_coffee_worksheet_investment_formulas}) has five cell references (\function{refs}{\texttt{Investment!E9:E11}}={\texttt{RC[-3]}, \texttt{RC[-2]}, \texttt{R5C2}, \texttt{RC[-1]}, \texttt{R4C2}}).
Therefore, it has five reference-based groups, namely \{\texttt{B4}\}, \{\texttt{B5}\}, \{\texttt{B9:B11}\}, \{\texttt{C9:C11}\}, and \{\texttt{D9:D11}\}.

The formula of the partitioned formula group \texttt{Department1!F4:F7} (Figure~\ref{fig:example_coffee_worksheet_department1_formulas}) has one area reference (\function{refs$_a$}{\texttt{Department1!F4:F7}}=\{\texttt{RC[-3]:RC[-1]}\}).
Therefore, the group refers to four reference-based groups, one for each of the four cells in the group: \{\texttt{B4:E4}\}, \{\texttt{B5:E5}\}, \{\texttt{B6:E6}\}, and \{\texttt{B7:E7}\}.
\end{example}


\change{
To allow for concise further processing of higher-level structures,} we then merge all overlapping reference-based groups that have the same orientation, i.e.
each cell is at most part of one reference-based group in vertical and one group in horizontal orientation.
\change{
The remaining non-empty cells that are contained neither in any partitioned formula group nor in any reference-based group are designated as \emph{non-blockable cells}, in preparation of the following Blocking step of the analysis.
}

\begin{definition}[Non-blockable cell]\label{def:non-blockable}
The function \function{non-blockable}{\param{cell}{c}} returns true, if \entity{c} is non-blockable:
$$\function{non-blockable}{c}=\\
\begin{cases}
		\text{true}  & \text{if } \function{type}{c}\neq \textit{empty} \wedge \{\nexists g \in G : c \in g\} \\
		\text{false} & \text{otherwise}.
	\end{cases}
$$
where $G=\function{partFormulaGroups}{\function{ws}{c}}\cup\function{referenceGroups}{\function{ws}{c}}$.
The function \function{non-blockables}{\param{worksheet}{w}} returns the set of all non-blockable cells contained in \entity{w} ($\function{non-blockables}{w}=\{c \in \function{cells}{w}: \function{non-blockable}{c}\}$).
\end{definition}

\begin{example}\label{ex:non-blockable}
The function \function{non-blockables}{\texttt{Department1}} returns \{\texttt{A1}; \texttt{B2}; \\ 
\texttt{A3}:\texttt{F3}; \texttt{A4}:\texttt{A8}\}, \function{non-blockables}{\texttt{Total}} returns \{\texttt{A1}; \texttt{B2}; \texttt{A3}:\texttt{F3}; \texttt{A4}:\texttt{A8}\}, and \\ \function{non-blockables}{\texttt{Investment}} returns \{\texttt{A1}:\texttt{A5}; \texttt{A7}:\texttt{A11}; \texttt{B8}:\texttt{E8}\}.
\end{example}

\subsection{Blocking}

In the blocking step, the formula groups and reference-based groups are aggregated to rectangular \emph{areas}.
\change{These areas are called \emph{blocks}, and each block} contains input cells, formula cells, and empty cells, but no header cells.

\begin{definition}[Area]\label{def:blockArea}
Function \function{area}{\param{\{cell\}}{C}, \param{worksheet}{w}} returns all cells contained in the rectangle spanned by the non-empty set of cells \entity{C} $=\{c_1,\dots,c_n\}$:
$$
\function{area}{C, w}=\{c \in \function{cells}{\function{ws}{c_1}} : x_1 \leq \function{x}{c} \leq x_2 \wedge y_1 \leq \function{y}{c} \leq y_2\}
$$
where 
$x_1 = \textit{min}\{\function{x}{c_1},\dotsc,\function{x}{c_n}\}$, 
$y_1 = \textit{min}\{\function{y}{c_1},\dotsc,\function{y}{c_n}\}$,\\
$x_2 = \textit{max}\{\function{x}{c_1},\dotsc,\function{x}{c_n}\}$, and
$y_2 = \textit{max}\{\function{y}{c_1},\dotsc,\function{y}{c_n}\}$.
\end{definition}

\begin{example}\label{ex:area}
\function{Area}{$\{\texttt{A3},\texttt{B4}, \texttt{C2}\}}$ computes $x_1=1$, $y_1=2$, $x_2=3$, and $y_2=4$ and returns as result \{\texttt{A2},\texttt{A3},\texttt{A4},\texttt{B2},\texttt{B3},\texttt{B4},\texttt{C2},\texttt{C3},\texttt{C4}\}.
\end{example}

\begin{definition}[Block]\label{def:block}
A non-empty set of cells \entity{C}~$=\{c_1,\dots,c_n\}$ forms a block if the area spanned by the cells \entity{C} contains no non-blockable cells:
$$\function{block}{C}=
\begin{cases}
		\text{true} & \text{if } \nexists c \in \function{area}{C}: \function{non-blockable}{c}\\
		\text{false} & \text{otherwise}.
	\end{cases}
$$
\end{definition}

\begin{example}\label{ex:Block}
For worksheet \texttt{Departement1} (Figure~\ref{fig:example_coffee_worksheet_department1_formulas}), \function{block}{$\{\texttt{B4},\texttt{D4}\}}$ returns \textit{true}, but \function{block}{$\{\texttt{A4},\texttt{D4}\}}$ returns \textit{false}, as cell \texttt{A4} is non-blockable.
\end{example}

\change{
Blocks are established by expansion operations that add \emph{block neighbors}, physically close groups, to an existing block.
A group is regarded as physically close to a block if there is at most one column or row between the block and the group which should be added.
We regard both, partitioned formula groups as well as reference-based groups, for block expansion.
}

\begin{definition}[Block Neighbors~\cite{iwpd/KochHW16}]\label{def:blockneighbor}
The function \function{neighbor}{\param{block}{b},\\ \param{group}{g}} returns \textit{true} if there is at most one row or column between \entity{b} and \entity{g}:
$$\function{neighbor}{b, g}=
\begin{cases}
		\text{true} & \text{if } \exists c \in g, \exists c' \in b: |x(c)-x(c')|\leq 2 \wedge y(c) = y(c') \\
		\text{true} & \text{if } \exists c \in g, \exists c' \in b: x(c)=x(c') \wedge |y(c) - y(c')| \leq 2 \\
		\text{false} & \text{otherwise}.
	\end{cases}
$$
\end{definition}

\change{Block creation for a worksheet follows the procedure of Algorithm~\ref{algo:BlockCreation}):}
First, the set of blocks is initialized (Line~\ref{algo:BlockCreation:init}) and the partitioned formula groups and the reference-based groups are computed (Lines~\ref{algo:BlockCreation:gp}-\ref{algo:BlockCreation:gr}).
The sets $G$ and $G'$ are both initialized with the union of the previously computed groups (Line~\ref{algo:BlockCreation:g}).
While $G'$ will be reduced in size during the computation of the blocks, $G$ never changes.
In the outer loop (Lines~\ref{algo:BlockCreation:outerLoopB}-\ref{algo:BlockCreation:outerLoopE}),
new blocks are created as long as there are groups which are not yet part of any block.
In the inner loop (Lines~\ref{algo:BlockCreation:innerLoopB}-\ref{algo:BlockCreation:innerLoopE}), groups are added to the blocks if they fulfil two criteria:
(1)~They must be neighboring to the block, and
(2)~the block and the group must form a valid block.
\change{Lastly, we return the inferred set of blocks for the given worksheet \entity{w}.}

\begin{algorithm}[ht]
	\caption{\textsc{Block Creation}}
	\label{algo:BlockCreation}
	\begin{algorithmic}[1]
		\Require worksheet w
		\Ensure set of blocks B
		
		\Procedure{blocks}{worksheet w}
			\State $B \leftarrow \emptyset$\label{algo:BlockCreation:init}
			\State $G_p \leftarrow \function{partFormulaGroups}{w}$\label{algo:BlockCreation:gp}
			\State $G_r \leftarrow \function{referenceGroups}{w}$\label{algo:BlockCreation:gr}
			\State $G,G' \leftarrow G_p \cup G_r$\label{algo:BlockCreation:g}
			\While{$\exists g\in G'$} \label{algo:BlockCreation:outerLoopB}
			  \State $b \leftarrow g$
				\State $G' \leftarrow G' \setminus g$
				\ForAll{$g' \in G$}\label{algo:BlockCreation:innerLoopB}
					\If{$\function{neighbor}{b,g'} \wedge \function{block}{b \cup g'}$}
						\State $b \leftarrow b \cup g'$
						\State $G' \leftarrow G' \setminus g'$
					\EndIf
				\EndFor\label{algo:BlockCreation:innerLoopE}
				\State $B \leftarrow B \cup \{b\}$
			\EndWhile\label{algo:BlockCreation:outerLoopE}
			\State \Return $B$
		\EndProcedure
	\end{algorithmic}
\end{algorithm}

\begin{figure}[ht]
    \centering
        \includegraphics[width=0.83\textwidth]{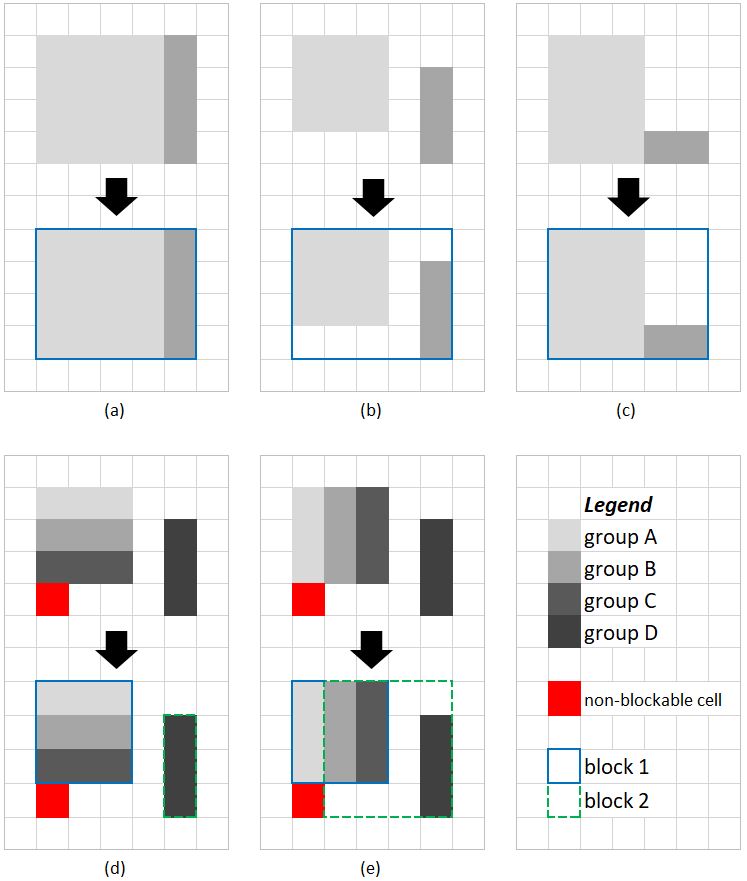}
    \caption{Block creation. Gray cells represent grouped cells, red cells represent non-blockable cells (i.e., non-empty cells that are not part of any group), and frames represent blocks.}\label{fig:block}
\end{figure}

Figure~\ref{fig:block} illustrates scenarios that might occur during the block creation.
In the sub-figures~(a), (b), (c), all groups can be merged to one large block.
Sub-figures~(d) and (e) additionally contain non-blockable cells. 
These cells prevent us from building a single large block.
In both cases, a second block (green hatched border) has to be created for the group to the right.
We allow groups to belong to several blocks. 
Hence, groups within the first block (blue solid border) might be added to the second block (green hatched border).
In sub-figure~(d), the groups of the first block have row-orientation.
Hence, adding any of these groups to the second block would violate the block criteria according to Def.~\ref{def:block}.
In contrast, all groups in sub-figure~(e) are column-oriented.
Therefore, some of the groups of the first block can also be added to the second one. 

\begin{example}\label{ex:blocking}
For the individual worksheets of our running example, the following blocks are computed:
\texttt{Department1!B4:F8}, \texttt{Total!B4:E8}, \texttt{Investment!\\B3:B5}, and \texttt{Investment!B9:E11}.
\end{example}

\subsection{Header Assignment}
In the third and final part of the structural analysis process, headers are assigned to blocks.
Headers are non-empty cells which are not part of a block, i.e., they are elements of the set of non-blockable cells.

The position of headers depends on the writing system:
The Left-To-Right (LTR) system, used in western countries, places headers left to and/or above blocks;
the Right-To-Left (RTL) system, used in Arabic countries, places headers right to and/or above blocks.
In the LTR system, the left-most cells have the lowest column index;
in the RTL system, the right-most cells have the lowest column index.
Therefore, we assume that headers of blocks have lower row- or column indices than the cells of the block.
In the following, we use the word \enquote*{left} to express that a cell has a lower column index than another cell, i.e., we focus on the LTR system, but the approach works similar in the RTL system.

Two types of cells can be located between a header cell $h$ and its block $b$: empty cells and other header cells. 
If a header cell is between $h$ and $b$, then $h$ is a higher level header, also called meta-header.
If there are no other header cells between $h$ and $b$, then $h$ is a low-level header.

\begin{figure}[htbp]
	\centering
		\includegraphics[width=0.80\textwidth]{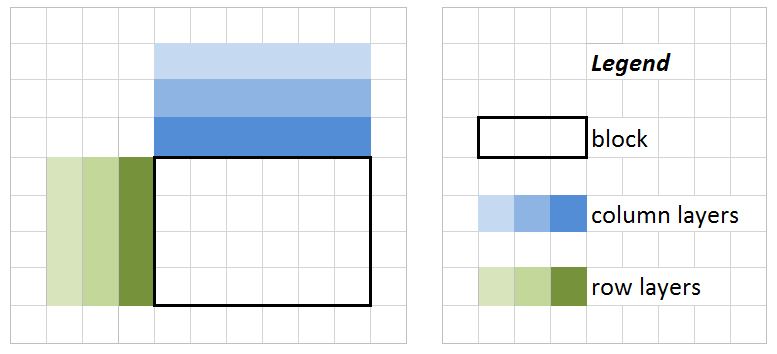}
	\caption{Column- and row layers for a block. The dark-shaded layers contain low-level headers; the light-colored layers contain higher-level headers.}
	\label{fig:BlockHeaderLayers}
\end{figure}

We call areas which contain headers layers.
There are row- and column layers.
Row layers are vertical groups which are located left to a block;
they have the same number of rows as the  block.
Column layers are horizontal groups which are located above a block;
they have the same number of columns as the underlying block.
Figure~\ref{fig:BlockHeaderLayers} illustrates the position and shape of layers.

We identify the column layers for a block $b$ by investigating the horizontal areas in the row above $b$.
If at least one of the cells of this area contains a non-blockable cell, a new layer has been detected.
In succession, we check the area above this layer.
We repeat this process until we reach the last row of potential headers or another block.
We analogously detect the row header layers.

\begin{definition}[Layers]\label{def:layers}
The function \function{columnLayers}{\param{block}{b}} returns the set of detected column header layers of block \entity{b}. Each header layer is described by the set of cells within its area. Likewise, the function \function{rowLayers}{\param{block}{b}} returns the set of detected column header layers of block \entity{b}.
\end{definition}

%

%

\begin{example}\label{ex:layers}
For our running example, we detect the following layers:
\begin{center}
	\begin{tabular}{l|ll}
	\textbf{Block}             & \textbf{Column layers}               & \textbf{Row layers}\\ \hline
	\texttt{Department1!B4:F8} & \texttt{B3:F3}, and \texttt{B2:F2}   & \texttt{A4:A8}\\
	\texttt{Total!B4:E8}       & \texttt{B3:E3} and \texttt{B2:E2}    & \texttt{A4:A8} \\
	\texttt{Investment!B3:B5}  & -                                    & \texttt{A3:A5} \\
	\texttt{Investment!B9:E11} & \texttt{B8:E8}                       & \texttt{A9:A11} \\
	\end{tabular}
\end{center}
The layers \texttt{Department1!B2:F2} and \texttt{Total!B2:E2} contain higher-level headers; the other layers contain low-level headers.
\end{example}

In the next step, we check the remaining non-blockable cells for being meta-headers:
If a non-blockable cell $c$ is left of a column layer or above a row layer, $c$ is a meta-header and has to be linked to the corresponding layer.
If $c$ is left to a column layer as well as above a row layer, we assign $c$ to the row layer, because row headers typically act as identifiers for single items of a set while column headers typically act as descriptive headers for different characteristics that are recorded for the set. 
Hence, $c$ is more likely to provide a common category for the underlying set of item-labels than a common category for the set of neighboring category labels.

\begin{example}\label{ex:meta-headers}
The following non-empty cells neither belong to a block nor to a layer: 
for the worksheets \texttt{Department1} and \texttt{Total}, the cells \texttt{A1} and \texttt{A3},
and for the worksheet \texttt{Investment}, the cells \texttt{A1}, \texttt{A2}, \texttt{A7}, and \texttt{A8}.
We are able to assign some of these cells to layers:
\begin{center}
	\begin{tabular}{l|l}
	\textbf{Cell }             & \textbf{Layer} \\ \hline
	\texttt{Department1!A3}    & \texttt{A4:A8} \\
	\texttt{Total!A3}          & \texttt{A4:A8} \\
	\texttt{Investment!A2}     & \texttt{A3:A5} \\
	\texttt{Investment!A8}     & \texttt{A9:A11} \\
	\end{tabular}
\end{center}
\end{example}

\subsection{Comparison with previous work}
We are not the first dealing with cell classification and header assignment for spreadsheets.
As mentioned in Section~\ref{sec:rw}, Abraham and Erwig~\cite{vlc/AbrahamE07} have identified header cells and core cells as part of the UCheck approach.
UCheck assigns one of four roles to each cell: 
(1) header (i.e. labels),
(2) footer (i.e. formula cells at the end of rows and/or columns which aggregate information), 
(3) core (i.e. data cells), and 
(4) filler (i.e. empty or particularly formatted cells which separate tables within spreadsheets). 
Thereby, UCheck uses different techniques to assign these roles, e.g. fence identification, content-based cell classification, and region-based cell classification.
Based on the results of the cell classification, UCheck assigns first-level headers to core and footer cells and higher-level headers to header cells.
Our analysis process is, in principle, based on the ideas of the cell classification and header assignment rules of UCheck. 
However, UCheck is not capable of identifying all header cells correctly. 
For example, UCheck fails to identify the headers for the quarters (cells \texttt{A4:A7} of worksheet \texttt{Total}) and the headers for the departments (cells \texttt{B2:E2} of worksheet \texttt{Total}) of the running example as headers.
Moreover, our analysis not only identifies cell roles, but also provides information about cohesive structures (groups/blocks) within a worksheet.

\section{Improved and New Spreadsheet Smell Detection Techniques}\label{sec:smellImprovements}
In this section, we demonstrate how the detected high-level structures can improve existing spreadsheet smell detection techniques, and how new smell detection techniques can be derived from the structure information. 

\subsection{Improved Smell Detection Techniques}\label{sec:smellUpdates}
\change{
The detected spreadsheet structures provide a number of opportunities to improve existing smell detection techniques.
Table~\ref{tab:refinement_suggestions} presents basic refinement ideas, for smells presented by Cunha \textit{et~al.}~\cite{CunhaFRS12}, and Hermans \textit{et~al.}~\cite{HermansPD12a,HermansPD12}.
}
\begin{table*}[thp]
	\centering
	\scriptsize
	\rowcolors{2}{gray!15}{white}
	\begin{tabular}{l|ccc|l} 
		\toprule
		\multicolumn{1}{c|}{\textbf{Name}}	& \multicolumn{1}{c}{\textbf{IS}} & \multicolumn{1}{c}{\textbf{FS}} & \multicolumn{1}{c}{\textbf{IWS}}	& \multicolumn{1}{|c}{\textbf{Refinement Suggestion}}\\
		\midrule
		Std. Deviation		& $\bullet$ & &	& Compare within group instead of column/row\\
		Empty Cell			& $\bullet$ & &	& Report connected vacant areas in blocks\\
		Pattern Finder		& $\bullet$ & &	& Compare within group instead of column/row\\
		String Distance 	& $\bullet$ & &	& Compare within group/block instead of sheet\\
		Ref. to Empty Cells	& $\bullet$ & &	& Highlight group instead of individual cells\\
		QFD					& $\bullet$ & &	& Compare within block instead of column/row\\
		Multiple Operations & & $\bullet$ &	& Report group instead of individual cells\\
		Multiple References	& & $\bullet$ &	& \makecell[l]{	Count group references instead of cell references\\
			Report group instead of individual cells}\\
		Cond. Complexity	& & $\bullet$ &	& Report group instead of individual cells\\
		Long Calc. Chain	& & $\bullet$ &	& \makecell[l]{	Count group references instead of cell references\\
			Report group instead of individual cells}\\
		Duplicated Formulas	& & $\bullet$ &	& \makecell[l]{	Detect duplicated formula groups\\
			Report group instead of individual cells}\\
		Inappr. Intimacy	& & & $\bullet$	& Count group references instead of cell references\\
		Feature Envy		& & & $\bullet$	& Count group references instead of cell references\\
		Middle Man			& & & $\bullet$	& Report group instead of individual cells\\
		Shotgun Surgery		& & & $\bullet$	& Count changing groups instead of formulas\\
		\bottomrule
	\end{tabular}
	\caption{Refinement suggestions for smells presented in the literature, categorized in input smells (\emph{IS}), formula smells (\emph{FS}), and inter-worksheet smells (\emph{IWS}).} 
	\label{tab:refinement_suggestions}
\end{table*}

\change{
Since similar smell detection techniques often can benefit from structure information in a similar way, we propose exemplary improvements for one representative of each smell group:
(1) for \textit{input smells} we improve the \emph{Pattern Finder} smell, 
(2) for \textit{formula smells} we refine the \emph{Long Calculation Chain} smell, and 
(3) for \textit{inter-worksheet smells} we improve the \emph{Feature Envy} smell.
For each investigated smell detection technique, we first discuss the original technique and its deficits. 
We then explain how to improve the smell detection process using structural information, and lastly discuss the benefits and drawbacks of the proposed improvements.
}

\subsubsection{Pattern Finder}\label{sec:improvedPatternFinder}
Cunha \etal \cite{CunhaFRS12} proposed the original Pattern Finder smell detection technique.
The smell detects cells that break a pattern that holds for the other cells of the same row or column, e.g., a constant in a row of formula cells or a string in a column of numerical values.
According to Cunha \etal \cite{CunhaFRS12}, this smell is detected by checking in windows of four cells if one of these cells has a different type than the other cells.
The authors provide an implementation of this technique in the \textit{FaultySheet Detective} tool \cite{AbreuCFMPS14a}. 
Examination of the tool\footnote{Version 1.1 from \url{http://ssaapp.di.uminho.pt/twiki/bin/view/Main/Software}} furthermore provided the following insights: 
	(1)~Patterns are detected in column orientation only.
	(2)~A cell's type refers to the type of its value, i.e., a cell with a constant numeric value within a series of formula cells which evaluate to a numeric value would not be indicated as smelly.
	(3)~A broken pattern can only be detected if no other cell in a 5-cell-distance above or below the cell features the same evaluated type.
	(4)~The smell is not detected for cells within the top or bottom five rows of a worksheet.

Pattern Finder attempts to establish patterns for entire columns of a worksheet. However, columns do not necessarily feature uniform content. 
For example, string cells are widely used to provide header information of a column. 
Similarly, cells at the bottom of a computation block might aggregate the data entered above or perform simple checks w.r.t. the validity of the above data.
This explains why the top and bottom 5 rows are excluded.
However, if a worksheet consists of several computation blocks, this workaround does not work.

We propose to focus smell detection on reference-based groups instead of generic sliding windows. 
This improves over the current detection process by allowing row-wise pattern detection and by extending the smell detection to the top and bottom five rows.
Moreover, we base the smell detection on non-evaluated cell types. 
This allows for the detection of instances where formulas and values are mixed in the same group. 

Algorithm~\ref{algo:detectPatternFinder} describes the updated Pattern Finder smell detection process. 
It detects reference-based groups whose cells feature more than one cell type. 
We iterate over the reference-based groups of the worksheet, checking for the smell (Lines~\ref{algo:detectPatternFinder:outerLoopB} to \ref{algo:detectPatternFinder:outerLoopE}).
In the inner loop (Lines~\ref{algo:detectPatternFinder:innerLoopB} to \ref{algo:detectPatternFinder:innerLoopE}),
we check whether group~G contains the smell:
If one of the cells has a different type, we add the group to the set of afflicted groups.
The algorithm returns this set as result.

\begin{algorithm}[ht]
	\caption{\textsc{DetectPatternFinder}}
	\label{algo:detectPatternFinder}
	\begin{algorithmic}[1]
		\Require worksheet w
		\Ensure reference groups afflicted by the Pattern Finder smell
		\Procedure{PatternFinderGroups}{worksheet w}\label{line:detectPatternFinder_function_countPatternFinderGroups}
			\State AfflictedGroups $= \emptyset$\label{algo:detectPatternFinder:init}
			\ForAll {G $\in$ \function{referenceGroups}{w}} \label{algo:detectPatternFinder:outerLoopB}
				\State Type $\leftarrow$ type of first cell in G 
				\Comment init
				\ForAll {c $\in$ \function{cells}{G}}\label{algo:detectPatternFinder:innerLoopB}
					\If {Type $\neq$ \function{type}{c}}	
						\State AfflictedGroups = AfflictedGroups $\cup \{G\}$
						\Break
					\EndIf
				\EndFor\label{algo:detectPatternFinder:innerLoopE}
			\EndFor\label{algo:detectPatternFinder:outerLoopE}
			\State \Return AfflictedGroups
		\EndProcedure

	\end{algorithmic}
\end{algorithm}

\begin{figure}[htbp]
	\centering
	\includegraphics[width=1.0\textwidth]{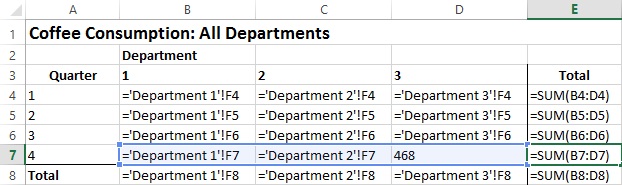}
	\caption{Example of Improved Pattern Finder smell detection technique}
	\label{fig:example_improvedPatternFinder}
\end{figure}

\begin{example}\label{ex:patternFinder}
The modified \texttt{Total} worksheet (Figure~\ref{fig:example_improvedPatternFinder}) illustrates the Improved Pattern Finder smell detection technique:
Cell \texttt{D4} has been changed from a formula to a fixed value.
The original Pattern Finder does not indicate \texttt{D4} as smelly, because every cell in column D evaluates to a number, but our improved Pattern Finder does.
Indicating  a constant within a group of formula cells as smelly helps users to detect formula cells which have been accidentally overwritten with constant values.
\end{example}


The focus on reference-based groups provides three key benefits in comparison to the original smell detection process: 
First, the algorithm compares cells within well-defined borders. This prevents the smell to be accidentally detected for column headers. Hence, the group-based approach reduces the number of false positives. 
Second, location limitations of the current detection approach do not apply for the updated algorithm. Groups in horizontal orientation and border areas of the worksheet are eligible for smell detection. 
Third, the updated algorithm only checks each reference-based group once, instead of checking every possible position of a sliding window, which makes the smell detection faster.

Setting the focus of the detection process on reference-based groups also introduces some drawbacks. First, smell detection is only applied to areas which are used as input values for calculations. Hence, the smell cannot be detected for areas containing output formulas, non-computational values, and labels. Further, the smell assumes that every referenced cell should contain a value at the time of smell detection. Consequently, blank spots which are reserved to be filled by a spreadsheet's user (as often used in form spreadsheets) will wrongly be indicated as smelly.
However, if all cells in the same reference group are empty, they feature the same type.
In such a case, no smell is reported.

\subsubsection{Long Calculation Chain}\label{sec:improvedLongCalculationChain}
Hermans \etal \cite{HermansPD12a} proposed the original Long Calculation Chain smell detection technique, which detects formulas referring to a long chain of formulas, because long calculation chains are difficult to follow and to understand.
This smell is detected by computing the maximum number of references which need to be followed when evaluating a formula.
If this number exceeds a certain threshold, the formula cell is indicated as smelly.

The main drawback of this smell detection technique  is its tendency to cause redundant calculations and detection notifications. 
Neighboring cells with the same R1C1 formula usually share the same preceding calculations. 
Hence, detectable issues for these cells can be traced back to one and the same general structural flaw. 
Computing the length of the calculation chain for each cell individually is inefficient. 
Moreover, 
each afflicted cell is reported individually. 
As a remedy, we propose to apply smell-detection on formula groups and inter-group dependencies instead of individual cells and cell references:

\begin{definition}\label{def:longestCalcChain}
The function \function{longestChain}{\param{partitioned formula group}{g}} calculates the longest chain of group \entity{g} as follows:
$$
\function{longestChain}{g} =\\
\begin{cases}
	\function{longestChain}{G} + 1 & \text{if } G \neq \emptyset\\
	1 & \text{else if } G_{r} \neq \emptyset\\
	0  &  \text{otherwise}
	
\end{cases}
$$
where 
$\function{longestChain}{G} = \text{max}\{l | \forall{g \in G}: l =\function{longestChain}{g}\}$, \\
\noindent
$G$=$\function{referredFormulaGroups}{g}$, and
$G_r$=$\function{referenceGroups}{g}$.
\end{definition}

We compute the length of the longest formula group chain of formula group~\entity{g} by adding 1 to the longest chain of the formula groups to which \entity{g} refers to. 
The chain of a formula group that has only references to input cells has a length of~1.
The chain of a formula group that has no references has a length of 0.

Like the original smell detection technique, the detection function of the group-based Long Calculation Chain smell returns a metric value. 
To decide on whether any given partitioned formula group is smelly, a threshold for the calculated metric is required. 
Groups whose calculated metric exceeds this threshold are indicated as smelly. Hermans~\etal proposed a threshold of 4 to indicate a small risk, and a threshold of 7 to indicate a high risk. 

\begin{example}\label{ex:longCalculatinoChain}
The \texttt{Investment} worksheet (Figure~\ref{fig:example_coffee_worksheet_investment_formulas}) illustrates the benefits of the Improved Long Calculation Chain:
Cells \texttt{E9}, \texttt{E10}, and \texttt{E11} are part of the partitioned formula group \texttt{E9:E11}. Each cell has a longest calculation chain of length 7. The specific references for each cell differ. However, the overall contextual structure of the calculation is shared among all cells. A specific reference chain path for cell \texttt{E9} is \texttt{Department1!B4} $\rightarrow$ \texttt{Department1!F4} $\rightarrow$ \texttt{Department1!F8} $\rightarrow$ \texttt{Total!B8} $\rightarrow$ \texttt{Total!E8} $\rightarrow$ \texttt{Investment!B3} $\rightarrow$ \texttt{Investment! B5} $\rightarrow$ \texttt{Investment!E9}. Similar paths can be reported for cells \texttt{E10} and \texttt{E11}. 
Alternatively, the Improved Long Calculation Chain reports only one calculation path for the entire group \texttt{Investment!E9:E11}. One possible group reference chain is 
\texttt{Department1!B4:E4} $\rightarrow$ \texttt{Department1!F4:F7} $\rightarrow$ \texttt{Department1!B8:F8} $\rightarrow$ \texttt{Total!B4:B8} $\rightarrow$ \texttt{Total!E4:E8} $\rightarrow$ \texttt{Investment!B3:B3} $\rightarrow$ \texttt{Investment!B5:B5} $\rightarrow$ \texttt{Investment!E9:E11}.
\end{example}

Checking for long calculation chains on a per-group basis has several benefits:
First, associated smells can be reported once per group instead of individually for each cell. 
This helps to reduce the number of reported smells. 
Second, per-group detection provides users with additional context to facilitate understanding of the overall calculation structure of the spreadsheet. 
A better mental model of a spreadsheet enables users to introduce well-considered changes. 
Third, group-wise smell detection implies that each reference is computed only once for an entire group of related formula cells. 
Hence, 
the detection approach may require substantially less individual references to be checked.

Setting the focus to group-based references 
introduces one key flaw: inconsistent groups. 
Inter-group references are not necessarily consistent for each individual cell within a group. 
Hence, the smell might be reported for a cell within a group even though this cell is not affected by the smell on a per-cell basis. 
As a remedy, only the partition of a formula group which is affected by the smell could be reported instead of the entire group. 
However, this would require per-cell detection of the smell in combination with per-group detection, neutralizing the calculation performance benefit of the improvement.
Moreover, although usual spreadsheet programs prohibit circular references on a per-cell reference basis, inconsistent group references might introduce circular reference paths in-between formula groups.
Such instances need to be handled correctly when calculating the metric's value.

\subsubsection{Feature Envy}\label{sec:improvedFeatureEnvy}
Hermans \etal \cite{HermansPD12} proposed the Feature Envy smell.
This smell detects worksheets which contain a large number of references to other worksheets.
It is difficult to follow many different relations to other worksheets when trying to understand or debug a spreadsheet.
Feature Envy reports worksheets for excesses in the number of individual connections to other worksheets. 
However, even a limited number of semantically different connections to other worksheets can render a worksheet difficult to comprehend and, thus, should be reported as smelly. 
Moreover, advanced tasks, e.g., elaborate data analysis, require a greater number of processing steps. 
Spreadsheet creators fulfil such tasks in two different ways:
(1)~They add more functionality into individual formulas, or add more formula cells to the worksheets.
(2)~They add new worksheets that refer to interim results. 
The first way increases either the complexity of the individual formula cells or the size of the worksheet; both consequences make a worksheet more difficult to understand.
Therefore, the second way becomes the preferred option at some point.
Consequently, we argue that a high number of semantically equivalent connections to the same worksheet should not necessarily be indicated as smelly. 
To allow for a more purposeful smell detection, we propose to base the smell detection on references of partitioned formula groups instead of individual formula connections.

Algorithm~\ref{algo:detectFeatureEnvy} describes the updated Feature Envy detection process. 
The function \function{ws}{\param{reference-based group}{g$_{r}$}} permits access to the worksheet of group $g_{r}$. 
The function \function{countFeatureEnvyConnections}{\param{worksheet}{w}} in Line~\ref{line:detectFeatureEnvy_function_countFeatureEnvyConnections} counts the number of total smell occurrences within worksheet $w$. 
It first initializes the variable $\textit{Count}$ with the value $0$. 
It then iterates the partitioned formula groups of worksheet $w$. 
For each group $g$, the function iterates the set of reference-based groups to which $g$ refers to. 
For each referred group $g_{\text{ref}}$, the function increments \textit{Count} if $g_{\text{ref}}$'s worksheet differs from worksheet $w$. 
Lastly, the function returns \textit{Count}, the number of group references to other worksheets.

\begin{algorithm}
	\caption{\textsc{DetectFeatureEnvy}}
	\label{algo:detectFeatureEnvy}
	\begin{algorithmic}[1]
		\Require worksheet w
		\Ensure number of connections from formula groups in w to other worksheets
		\Procedure{countFeatureEnvyConnections}{worksheet w}
		\label{line:detectFeatureEnvy_function_countFeatureEnvyConnections}
		\State Count $\leftarrow$ 0
		\ForAll {g $\in$ \function{partFormulaGroups}{w}} \label{algo:detectFeatureEnvy:formulaGroups}
			\ForAll {g$_{\text{ref}}$ $\in$ \function{referenceGroups}{g}}
				\If {\function{ws}{g$_{\text{ref}}$} $\neq$ w}
					\State Count $\leftarrow$ Count + 1
				\EndIf
			\EndFor
		\EndFor
		\State \Return Count
		\EndProcedure
	\end{algorithmic}
\end{algorithm}

\begin{example}\label{ex:featureEnvy}
The cells in the area \texttt{B4:D8} of worksheet \texttt{Total} (Figure~\ref{fig:example_coffee_worksheet_total_formulas}) feature 12 references to other worksheets. However, the cells in each column can be grouped into the partitioned formula groups \texttt{B4:B8}, \texttt{C4:C8}, and \texttt{D4:D8}.
Hence, the Improved Feature Envy only reports three inter-worksheet connections.
\end{example}

The detection function of the group-based Feature Envy smell returns a value. 
Worksheets whose calculated metric value exceeds a certain threshold are indicated as smelly. Hermans~\etal proposed a threshold of 3 to indicate a small risk, and a threshold of 7 to indicate a high risk.


The proposed improvement offers two main benefits: 
First, applying group connections for the calculation of the detection metric provides users with more meaningful feedback in regard to the overall quality of the connection structure of the spreadsheet. This supports users in making high-level structural improvements, eliminating the root cause of indicated smells instead of alleviating its effects.
Second, group-based smell detection only requires to check for inter-worksheet connections of each partitioned formula group, instead of checking the connections of all cells of a worksheet.
Hence, 
this detection approach may require substantially less individual references to be checked.

The drawback of using group connections is a potential loss of information. While a high number of semantically similar inter-worksheet connections might be a necessary design, 
reporting the circumstance might still offer an opportunity for improvement. As a remedy, the cardinality of processed group connections might be introduced into the detection metric, be reported to the user as contextual information, or both.

\subsection{New smell detection techniques}\label{sec:novelSmells}  

We elaborated different approaches to formulate new smell detection methods that are based on the structural information.
From this list of fundamental smell ideas, we present three smells that showcase utilization of different aspects of the available structure information:
The Overburdened Worksheet smell indicates that a worksheet contains too much functionality. 
The Inconsistent Formula Group Reference smell signals inconsistencies occurring within the group-resolving step of the analysis process.
The Missing Header smell indicates gaps in the headers of a block. 
For each of the introduced smells, we first outline its basic  concept. We then present the smell detection process and provide an example.
Lastly, we highlight benefits and possible limitations of the new smell, and explain its significance related to the overall quality of a spreadsheet.
For doing so, we will refer to the ISO/IEC 25010:2011 International Standard for System and Software Quality Models~\cite{ISO25010} and the quality model~\cite{CunhaFPS12} for spreadsheets that is based on a predecessor of this standard.

\subsubsection{Overburdened Worksheet}\label{sec:overburdenedWorksheet}

Each part of a spreadsheet serves a specific purpose: 
a reference group provides a set of common input data for further calculations;
a formula group performs a calculation on sets of input data;
a block distinguishes functionally enclosed areas of a worksheet. 
The Overburdened Worksheet smell indicates worksheets which feature an excessive number of any structure type, e.g., blocks:

\begin{definition}[Overburened Worksheet]\label{def:overburdenedWorksheet}
The function \function{overburdenedWorksheet}{\param{worksheet}{w}} returns the detection metric for the smell:
$$
\function{overburdenedWorksheet}{w} = |\function{blocks}{w}|.
$$
\end{definition}

To decide on whether any given worksheet is smelly, a threshold for the calculated value is required. 
Worksheets whose calculated value exceeds this threshold are indicated as smelly.

\begin{example}\label{ex:overburendedWorksheet}
The \texttt{Investment} worksheet (Figure \ref{fig:example_coffee_worksheet_investment_formulas}) 
has two calculation blocks: \texttt{B3:B5} and \texttt{B9:E11}.
If we set the threshold of this smell to the extremely low level of two, the worksheet would be indicated as smelly.
\end{example}

As worksheet \textit{Investment} contains a minimal example, it remains comprehensible despite featuring multiple calculation blocks. However, in general, multiple blocks in a worksheet indicate a suboptimal spreadsheet structure, which can be easily resolved by moving some blocks to a new worksheet.

The provided detection function utilizes the number of calculation blocks per spreadsheet as significance metric. 
However, other structure information may be employed, as well. 
We included the number of formula groups per worksheet as an additional metric in our evaluation. 
Another possibility would be to count the reference-based groups of a worksheet, or the number of intra-worksheet connections in-between formula groups.
When relying on basic spreadsheet information, the number of cells or the number of formulas might also be employed to indicate an overburdened worksheet.

The Overburdened Worksheet smell provides a natural counter-balance to existing inter-worksheet smells indicating worksheets that refer too abundantly to other worksheets.
The goal for an overall, optimal spreadsheet structure is then a balanced partition of the required functionality over a number of worksheets which neither overburdens any individual sheet nor renders any sheet overly reliant on inter-worksheet connections. 

The quality and success of the Overburdened Worksheet's smell detection process highly depends on the success of the previous structural analysis process. 
Inconclusive structural analysis might result in an excessive number of small structures. 
In such a case, the size-based metrics provide misleading information. Smell metrics which combine the quantity of structures with their respective size might lessen the influence of ambiguous structure analysis.

While an overburdened worksheet does not influence the functionality and security of a spreadsheet, it  influences the maintainability (in particular the subcatetory analyzability, see ISO/IEC 25010:2011 standard~\cite{ISO25010}) and usability (subcategory understandability):
A spreadsheet that has a clear modular structure with worksheets as modules is easier to understand and maintain than a spreadsheet that contains all information in a single worksheet.

\subsubsection{Inconsistent Formula Group Reference}\label{sec:inconsistentFormulaGroupReference}


The Inconsistent Formula Group Reference smell highlights an inconsistency that becomes apparent during the structural analysis process.
Common spreadsheet programs already point out inconsistencies regarding the formulas of groups of related cells. 
Inconsistent Formula Group Reference points out inconsistencies regarding references to individual cells of such groups.

More elaborate worksheets depend on reference chains, linking sequential calculations. 
Structural analysis enables tracking of references in between formula cells, as well as references in between formula groups. 
However, references in between formula groups are not always concise. 
For example, one group might refer only to a subset of the cells of another group.
Inconsistent Formula Group Reference points out those instances.

\begin{definition}[Inconsistent group reference]\label{def:inconsistentGroupReference}
	The function \\ \function{inconsGroupRef}{\param{partitioned formula group}{g}, \param{partitioned formula group}{g'}} identifies inconsistent group references:
	$$
	\function{inconsGroupRef}{g,g'} =
	\begin{cases}
	\text{true} & \text{if} ~ \nexists~g_r \in \function{referenceGroups}{g} : g_r \equiv g' \\
	& \wedge~ \exists~ g_r \in \function{referenceGroups}{g} : g_r \cap g' \neq \emptyset\\
	\text{false} & \text{otherwise}.
	\end{cases}
	$$
\end{definition}

\begin{example}\label{ex:inconsistentFormulaGroupReference}
Our running example in Figure~\ref{fig:example_coffee_formulas} contains an occurrence of the Inconsistent Formula Group Reference smell. In the \texttt{Investment} worksheet, cell~\texttt{B3} creates a formula group that refers to the single cell \texttt{E8} of the \texttt{Total} worksheet. However, the cell \texttt{Total!E8} is part of the formula group \texttt{Total!E4:E8}. Hence, \texttt{Investment!B3:B3} inconsistently refers to \texttt{Total!E4:E8}.  
\end{example}

The smell points to inconsistencies within reference chains in between formula groups. 
Such inconsistencies may be introduced during the creation or expansion of the spreadsheet, e.g., a newly created set of calculations mistakenly refers to only a part of preceding formulas, or an inner part of a calculation chain is expanded, but successive calculations are not updated accordingly.

As demonstrated in the example, referring only to a part of a formula group may not always indicate an error, but be the intended behavior. 
However, even in those cases, a spreadsheet may be restructured to remove the inconsistency. 
Moreover, inconsistency detection depends on the success of the previous structural analysis process. 
Incorrect grouping, poor partitioning of formula groups, or resolving of group references might lead to false positive detection.

A spreadsheet with inconsistent references is difficult to analyze.
Inconsistent references might point to faults caused by expanding a spreadsheet.
While some parts of a spreadsheet are updated, others might be forgotten to be updated.
When the forgotten updates lead to errors, the spreadsheet does not provide the intended functionality.
Hence, the Inconsistent Formula Group Reference influences the overall quality of a spreadsheet with respect to its analyzability (subcategory of maintainability) and to a certain extent to its functionality (see ISO/IEC 25010:2011 standard~\cite{ISO25010}).

\subsubsection{Missing Header}\label{sec:missingHeader}

Headers are not always provided for each column and/or row of a block.
This results in empty spots within the header layers of affected blocks.
The Missing Header smell reports cases of such vacant spots.

\begin{definition}[Missing headers]\label{def:missingHeaders}
	The function \function{missingHeaders}{\param{block}{b}} returns the set of missing header cells of block \entity{b}:
	$$
	\function{missingHeaders}{b} = \{~c \in ( \bigcup_{l \in L_\text{col}} l ) \cup ( \bigcup_{l \in L_\text{row}} l )  :  \function{type}{c}=\text{\enquote*{empty}}~\}
	$$
	where $L_\text{col}=\function{columnLayers}{b}$ and $L_\text{row}=\function{rowLayers}{b}$.
\end{definition}

\begin{figure}[htbp]
	\centering
	\includegraphics[width=1.0\textwidth]{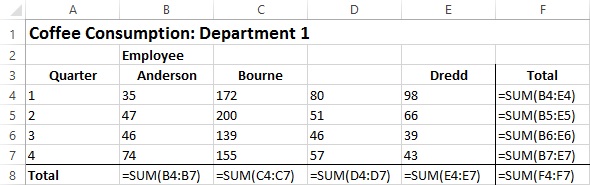}
	\caption{Example of Missing Header smell}
	\label{fig:example_missingHeader}
\end{figure}

\begin{example}\label{ex:missingHeader}
Figure~\ref{fig:example_missingHeader} illustrates the Missing Header smell. 
It depicts a revised version of the \texttt{Department1} worksheet of our running example. 
For demonstration purposes, we have removed the label of cell \texttt{D3}.
Structural analysis detects a block in area \texttt{B4:F8}. 
Column-headers for this block are available in row 3. 
However, one spot in the header layer of the block, cell D3, is vacant. 
\end{example}

The Missing Header smell gives feedback about the quality of non-calculation parts of worksheets.
Missing headers impair comprehensibility of worksheets, as calculation relations do not necessarily provide contextual information.

Inference of headers and header layers is directly dependent on the preceding block detection results. 
Hence, header detection carries on the same limitations as affected the previous analysis steps. 
For example, the current blocking approach does not compute blocks for tables that only collect data, but do not process it. 
Thus, no headers can currently be inferred for such tables.
Another drawback of the proposed analysis method is a high likelihood of false positives in higher-order header layers.
Items of such layers usually provide contextual information to underlying header layers; they are, therefore, usually not completely filled in. 
This is an intended behavior. Nevertheless, the current definition would count such instances as missing headers.

Moreover, conflicts may occur, whereby meta-headers can be assigned to both, underlying column- and row header layers. 
Following the current header assignment process, such conflicts are resolved by a static default decision. 
To attain more reliable results for higher-order headers, structural analysis would benefit from a more elaborate approach to correctly decide ties.

Badly or even undocumented spreadsheets are obviously difficult to understand.
Hence, missing headers influence the overall quality of a spreadsheet by a reduced understandability (subcategory of the usability characteristic in the ISO/IEC 25010:2011 standard~\cite{ISO25010}).

\section{Empirical Evaluation}\label{sec:eval}

In this section, we evaluate the performance of the improved and new smell detection techniques. 
We first outline our study design.
We then present the details and results and lastly, we discuss the presented results.

\subsection{Study Design}\label{sec:studyDesign}


\textbf{Study Rationale.} 
The rationale of this study is to evaluate whether structural information improves the detection of spreadsheet smells.
The improvement potential for smells varies, based on the source smell and on how structure information can be applied. 
In general, we assume that the improvement of the detection techniques results in a reduction of false positives, and limits the number of redundant smell detections.
Newly introduced smell detection techniques are expected to perform similar to already established ones.

The performance analysis of the underlying structural detection process is out of the focus of this empirical evaluation.
We refer the interested reader to our IWPD paper~\cite{iwpd/KochHW16} for a detailed evaluation of the detection performance for blocks, groups, and headers.

\textbf{Objective/Units of study \& Context.} 
The objective of this particular evaluation is to investigate the detection performances of the improved and new spreadsheet smell detection techniques.
The context of our study is the EUSES corpus, a publicly available collection of spreadsheets.
Units of analysis are the sets of spreadsheet smells: original smells, improved smells, and new smells.
For each of these sets, we record the respective detection metrics when applied to spreadsheets of the EUSES~\cite{sigsoft/FisherR05} corpus.



\textbf{Research questions.}
\begin{enumerate}
	\item \rqone
	\item \rqtwo
\end{enumerate}


\textbf{Concepts \& Measures.}
As established by previous research \cite{HermansPD12a}\cite{HermansPD12}, we use detection metrics of smells as basis of analysis and comparison; to determine whether a given entity is smelly, every smell detection technique calculates a metric value for the entity.
\change{
The target entities for baseline techniques are either cells or worksheets, whereas entities for the improved and new techniques are either groups, blocks or worksheets.
To allow for comparison of the approaches, we thus regard a cell as well as a group and a block as individual entities of a spreadsheet which a user has to check, if indicated as smelly.
Metrics that are recorded per worksheet are directly comparable.} 
Following common practice, we aggregate these metrics into quartile plots.
Quartile plots aggregate the results of individual entities, illustrating the percentage \texttt{x} of the analyzed entities that feature a metric value of \texttt{y} or lower.
To incorporate a wider range of metric values, we use a logarithmic scale for the y-axis.
This allows us to compare our results with previous work.


\textbf{Data Collection.}
Data collection is based on our own test implementation, \fritz\footnote{available at \url{http://spreadsheets.ist.tugraz.at/index.php/software/}}. 
The \enquote*{evaluation} command of \fritz\ analyzes a supplied spreadsheet corpus and writes a selectable set of calculated metrics to CSV files for further processing.
The evaluation uses files of the EUSES~\cite{sigsoft/FisherR05} spreadsheet corpus.
The corpus can be downloaded from the tera{-}PROMISE Repository\footnote{\url{http://openscience.us/repo/spreadsheet/euses.html}, last visited 2017-01-31}. 
It contains 4490 files in 11 categories.
Not all files of this corpus are fit for evaluation with \fritz.
In a preprocessing step, we exclude files which 
(1)~are not readable by external library components used by the evaluation tool, 
(2)~are not processable due to limitations of the evaluation tool, or 
(3)~do not contain any formulas.
This preprocessing operation is provided by \fritz, using the command \enquote*{preprocess}.
This command supports different filtering options, e.g., 
the \enquote*{complete} option which applies all the previously mentioned filter criteria. 
The resulting filtered EUSES corpus consists of 1735 files in 10 categories.
We then run the automatic \enquote*{evaluation} command offered by \fritz, using the evaluation option \enquote*{SMELLS\_COMPLETE}.
\fritz\ has a 5~minute timeout limit per file. 
Three files (\textit{personal/FindFunction.xls}, \textit{inventory/in\_emit99.xls}, and \textit{grades/PregnancyDiet.XLS}) exceed this limit.
The stated results refer to the 1732~files that are fit for evaluation and do not cause a timeout.


\textbf{Data Analysis.}
In order to compare our improvements to the original smell detection techniques, we have to introduce a common baseline.
To establish this baseline, we collect data on the detection performance of our own re-implementations of the basic smell detection approaches.
For our data analysis, we first compare the new baseline with the results of the smells' original authors where available.
We then compare the collected data for the improved smell detection techniques to our baseline results.
For our set of novel smells, we analyze the smell detection performance following the established analysis approach for spreadsheet smells and provide general remarks.

\textbf{Case \& Data Selection.}
	The general case of the study is provided by the spreadsheets of the EUSES corpus. 
	Units of analysis within the study are the sets of smells: baseline, improved, new.
	The data resulting of analyzing all eligible spreadsheets within the corpus is used for each analysis unit. 
	Eligible spreadsheets are determined in a preprocessing step, using the \fritz tool.

\textbf{Replication.}
	The focus of the present study is to build up on, rather than to replicate existing work. 
	However, to allow for comparison, we had to replicate some parts of related publications, as the tools and data of those studies are no longer available.
	In order to guarantee that our work is replicable, we provide references to the used dataset and tools.



\subsection{Smell Detection Improvements}
Since the tools that were used for evaluating the original smell detection process are either not publicly available or not designed to support automatic evaluation using a spreadsheet corpus, we implemented the baseline smell detection techniques in our own evaluation tool, \fritz.
For each improved smell, we first compare our baseline implementation with the smell's original evaluation results (using it's original evaluation dataset).
We then compare this baseline implementation with our improved variant, using the EUSES corpus as dataset.
\change{
As we want to highlight the reduction in total detections caused by avoiding redundancies, the improved versions of the \emph{Long Calculation Chain} and \emph{Feature Envy} smells use the same thresholds for detection as the baseline techniques.
}

\subsubsection{Pattern Finder} \label{sec:pattern_finder}
Cunha~\etal implemented this technique in the FaultySheet Detective tool, and evaluated it using 180 selected spreadsheets of the EUSES corpus.
To enable comparison, we reproduced the subset of spreadsheets after consultation of the authors. 
\change{
Since the FaultySheet detective tool does not support a batch-mode analysis, the manual execution of the tool for each spreadsheet and evaluation of the results is very time consuming.
Hence, we chose to limit the comparison of our implementation with the author's original results using the FaultySheet Detective\footnote{Version 1.1 from \url{http://ssaapp.di.uminho.pt/twiki/bin/view/Main/Software}} to the category \enquote{homework} (16 spreadsheets).
However, we provide corpus archive as download\footnote{\url{http://spreadsheets.ist.tugraz.at/wp-content/uploads/EUSES_small.zip}}, to allow for validation of our results. }
Table~\ref{table_smellimprovements_patternfinder} illustrates the analysis results using the following metrics:
\begin{itemize}
	\item \texttt{FaultySheet}: smell instances detected by FaultySheet Detective.
	\item \texttt{Relevant}: amount of \texttt{FaultySheet} detections we regard as relevant, based on manual inspection. 
	A relevant detection indicates a cell which discerns from the obvious intention of the spreadsheet's author (e.g., a number instead of a date).
	Other detections (e.g., labels and descriptions within a table) are not regarded as relevant.
	\item \texttt{Cols}: smell instances detected by \fritz using column-oriented windows. This corresponds to FaultySheet Detective's detection approach. 
	\item \texttt{Rows}: smell instances detected by \fritz using row-oriented windows.
\end{itemize}
\change{\texttt{FaultySheet} and \texttt{Relevant} numbers result from manual inspections of the resulting sheets by one researcher.}
For each metric, we provide the total number of detections for all analyzed worksheets (\emph{Cells Total}), as well as the average and median number of detections per worksheet (\emph{Cells Average} and \emph{Cells Median}).
Moreover, we state the number and percentage of worksheets that feature any detection (\emph{Worksheets ($>$0)} and \emph{\% Worksheets ($>$0)}), as well as the average number of detections for these worksheets (\emph{Cells Average ($>$0)}).

\begin{table*}[thp]
	\centering
	\scriptsize
	\rowcolors{2}{gray!15}{white}
	\begin{tabular}{l|rrrr} 
		\toprule
		\multicolumn{1}{c|}{\textbf{Metric}}	& \multicolumn{1}{c}{\textbf{FaultySheet}} & \multicolumn{1}{c}{\textbf{Relevant}} & \multicolumn{1}{c}{\textbf{Cols}}	& \multicolumn{1}{c}{\textbf{Rows}}\\
		\midrule
		Cells Total			& \tbldata{\numprint{180}}		& \tbldata{\numprint{20}}		& \tbldata{\numprint{181}}		& \tbldata{\numprint{129}} \\
Cells Average		& \tbldata{\numprint{6,6}}		& \tbldata{\numprint{0,7}}		& \tbldata{\numprint{6,6}}		& \tbldata{\numprint{4,8}} \\
Cells Median		& \tbldata{\numprint{0,5}}		& \tbldata{\numprint{0}}		& \tbldata{\numprint{1}} 		& \tbldata{\numprint{0}} \\
Worksheets ($>$0)	& \tbldata{\numprint{14}}		& \tbldata{\numprint{4}}		& \tbldata{\numprint{15}} 		& \tbldata{\numprint{9}} \\
\% Worksheets ($>$0)& \tbldata{\numprint{50}}{\%}	& \tbldata{\numprint{14}}{\%}	& \tbldata{\numprint{54}}{\%}	& \tbldata{\numprint{32}}{\%} \\
Cells Average ($>$0)& \tbldata{\numprint{12,9}}	& \tbldata{\numprint{5,0}}		& \tbldata{\numprint{12,1}} 	& \tbldata{\numprint{14,3}} \\
		\bottomrule
	\end{tabular}
	\caption[Pattern Finder smell detection metrics.]{Pattern Finder detection metrics based on the homework folder of Cunha~\textit{et al.}'s evaluation set. Cell count average and median are calculated on a per-worksheet basis.}
\label{table_smellimprovements_patternfinder}
\end{table*}

Our recorded detection numbers for the FaultySheet tool diverge from the numbers stated by the smell's authors \cite{CunhaFRS12}. 
Cunha~\etal reported 58 detected Pattern Finder smells; 56 of these they categorized as \enquote{no smells}, leaving two genuine smell detections.
In contrast, we counted 180 smelly cells as detected by the FaultySheet Detective tool.
Manual inspection categorized 20 of these as relevant detections, \change{for example the number 38412 in a column labelled \enquote{Target date for next steps} that otherwise contained proper date values.}
In comparison, our analysis tool managed to find the same 180 smell instances using column-oriented windows. 
\change{
Hence, the baseline implementation of the technique adequately reproduces the performance of the original approach.
}
One additional instance was detected due to technical specifics of a utilized library component. 
When using row-based windows, \fritz detected 129 smelly cells.

\begin{figure}[htbp]
	\centering
	\includegraphics[clip, width=1.0\textwidth, trim=2cm 9.5cm 1.5cm 9cm]{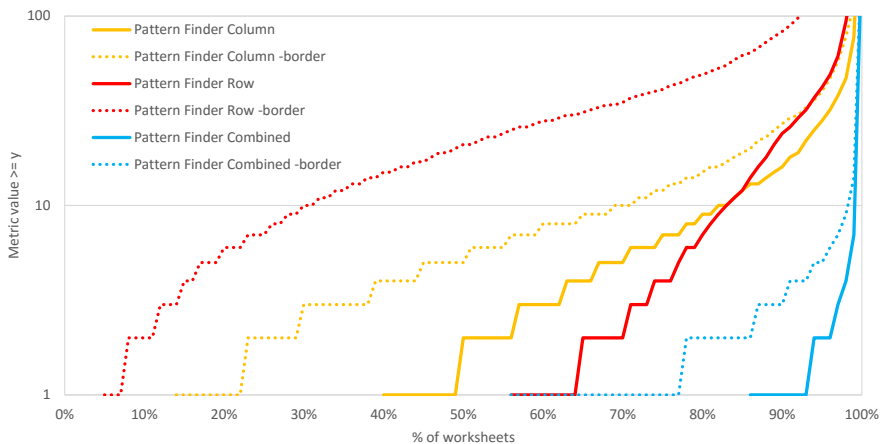}
	\caption{Quartile plot of various interpretations of the original Pattern Finder smell, evaluated on EUSES.
	See Section~\ref{sec:studyDesign} Concepts \& Measures for a description of quartile plots.}
	\label{fig:figure_eval_patternFinder_baseline}
\end{figure}

\change{
We thus proved the adequacy of our implementation when compared to the original tool.
However, during this evaluation we also revealed a significant shortcoming in terms of relevant detections.
We assume that specific implementation details are to blame for these shortcomings.
To check how specific implementation choices influence the results, we devised a number of different interpretations of the base technique, and evaluated the approach on the EUSES corpus.}
Figure~\ref{fig:figure_eval_patternFinder_baseline}~illustrates the results of the evaluation of various interpretations of the Pattern Finder smell as implemented in \fritz. 
\emph{Pattern Finder Column} and \emph{Pattern Finder Row} only use detection windows in the respective orientation and exclude detections in the first and last five columns and rows.
The \emph{Pattern Finder Column -border} and \emph{Pattern Finder Row -border} metrics work the same, but suspend the border restrictions.
The \emph{Pattern Finder Combined} metric reports cases where a cell is indicated as smelly by both, a horizontal and a vertical detection window, and excludes detections in border areas.
The \emph{Pattern Finder Combined -border} metric works the same way, but also allows detections in the border areas.
Outliers that exceed metric values of 100 are not depicted.

\emph{Pattern Finder Column} detects more smell instances than its row-based counterpart. 
The detection numbers of \emph{Pattern Finder Combined} is significantly lower than both, indicating that only a limited overlap exists between column-based and row-based smell detections.
When allowed to detect smells in border areas (\emph{-border}), the detection rates of all approaches increase noticeably. 
The rise is especially significant for the \emph{Pattern Finder Row -border} metric, where at least one smell instance is detected for more than 90\,\% of worksheets.
Borders of worksheets usually contain a high number of string cells that are used as descriptive headers and footers.
Hence, we suspect that the additional detections in border areas include a high number of such cells, and can thus be regarded as false positive detections.
\emph{Pattern Finder Combined -border} also registers a moderate increase in its detection rate in comparison to the results of it's border-excluding counterpart.
Nevertheless, it identifies fewer smelly cells than the comparable row- and column-based approaches.
We argue that the combination of row-based and column-based detection windows reduces the likelihood of false positive detections by counteracting the tendency of spurious detections in border areas.
We consequently regard \emph{Pattern Finder Combined -border} as the most comprehensible of the analyzed approaches, and add this measure as additional baseline for comparison with our \emph{Improved Pattern Finder} approach.

\begin{figure}[htbp]
	\centering
	\includegraphics[clip, width=1.0\textwidth, trim=2cm 9.5cm 1.5cm 9cm]{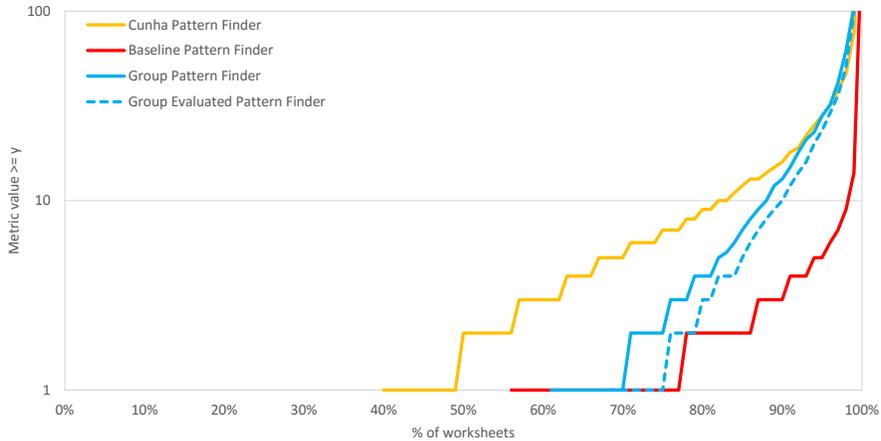}
	\caption{Quartile plot comparing the original Pattern Finder, our selected baseline variant, and two improved smell variants on the EUSES dataset.}
	\label{fig:figure_eval_patternFinder_comparison}
\end{figure}

Figure~\ref{fig:figure_eval_patternFinder_comparison} compares two interpretations of our improved smell detection approach with the original technique, \emph{Cunha Pattern Finder}, and the selected baseline variant, \emph{Combined Pattern Finder -border}.
The metric \emph{Group Pattern Finder} illustrates the result of our implementation based on Algorithm~\ref{algo:detectPatternFinder}.
The \emph{Group Evaluated Pattern Finder} first evaluates formula cells and uses the evaluated cell types for comparison with other cells.
In total, \emph{Cunha Pattern Finder} detected \numprint{45010} smelly cells (\numprint{7,4} per worksheet), and \emph{Pattern Finder Combined -border} used as baseline, identified \numprint{8003} smelly cells (\numprint{1,3} detections per worksheet).
In comparison, \emph{Group Pattern Finder} counted \numprint{49667} group detections, (\numprint{8,2} per worksheet), and \emph{Group Evaluated Pattern Finder} counted \numprint{44453} group detections (\numprint{7,3} per worksheet).

As expected, \emph{Cunha Pattern Finder} reports a significantly larger number of smell detections than the other approaches, detecting at least one smelly cell within 40\,\% of analyzed worksheets.
\emph{Group Evaluated Pattern Finder}'s result is similar to the chosen baseline.
Both evaluate formula cells before comparing the types, hence both approaches are likely to detect similar cases of the smell.
The higher number of detections of the \emph{Group Evaluated Pattern Finder} is likely due to cases where a pattern is broken only in one orientation, but not the other.
The results of \emph{Group Pattern Finder} follow the same overall trend as those of \emph{Group Evaluated Pattern Finder}.
Due to cases whereby the evaluation of formula cells conceals otherwise mismatching cell types, the number of individual detections is higher. 
As such instances might indicate genuine deficits, we suggest using \emph{Group Pattern Finder} over the \emph{Group Evaluated Pattern Finder approach}.

\subsubsection{Long Calculation Chain}
\begin{figure}[htbp]
	\centering
	\includegraphics[clip, width=1.0\textwidth, trim=2cm 9.5cm 1.5cm 9cm]{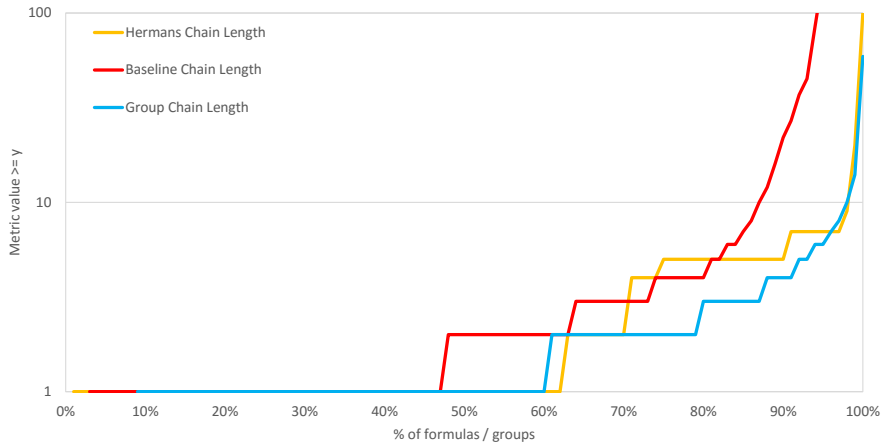}
	\caption{Quartile plot of Long Calculation Chain metrics}
	\label{fig:figure_eval_longCalculationChain_comparison}
\end{figure}

Figure~\ref{fig:figure_eval_longCalculationChain_comparison} compares our evaluation results with the author's initial results.
\emph{Hermans Chain Length} indicates the evaluation result as published by Hermans~\etal \cite{HermansPD12a}.
\emph{Baseline Chain Length} is the result of our interpretation of the smell using \fritz.
\emph{Group Chain Length} refers to the results of the improved smell detection version, as described in Section~\ref{sec:improvedLongCalculationChain}.


\emph{Baseline Chain Length} computes an average chain length of \numprint{24.5}, while \emph{Group Chain Length} computes an average chain length of \numprint{2.1}. 
When applying the threshold of 7 for a smell detection, the baseline approach identifies \numprint{84031} formula cells as smelly, whereas the group-based approach identifies only \numprint{4879} groups  as smelly.
\change{
Consultation of the smell's original authors revealed that detailed numbers for \emph{Hermans Chain Length} are no longer available. 
The presented numbers are thus extracted from the result plots given in \cite{HermansPD12a}.
}

When comparing significant features of the graph, our baseline result indicates a higher proportion of calculation chains as smelly than was indicated by the original results. 
Moreover, \emph{Baseline Chain Length} features a greater proportion of formulas that feature chain lengths significantly longer than 10 than both the original and group-based smell versions.
These deviations might be caused by specific design choices regarding the compared smell implementations, or by distinctions in the evaluation datasets introduced by pre-processing operations.
The graph of \emph{Group Chain Length} follows a similar progression as the graph of the baseline implementation. 
The number of individual detections, however, is noticeably lower. 
This is the expected result, as the same structural flaws are detected by both approaches.
Nevertheless, the baseline implementation has a higher individual metric number due to redundant smell detections.
In terms of detection rates, the baseline approach results in a substantially larger number of individual detections than the group-based approach. 
\emph{Group Chain Length}, therefore, offers a more concise way of communicating these flaws.


\subsubsection{Feature Envy}
\begin{figure}[htbp]
	\centering
	\includegraphics[clip, width=1.0\textwidth, trim=2cm 9.5cm 1.5cm 9cm]{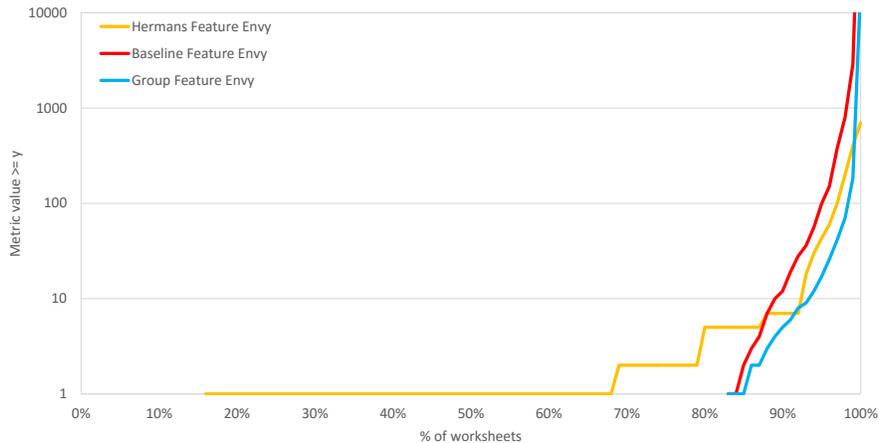}
	\caption{Quartile plot of Feature Envy metrics}
	\label{fig:figure_eval_featureEnvy_comparison}
\end{figure}

Figure~\ref{fig:figure_eval_featureEnvy_comparison} compares the results of the improved Feature Envy detection metric to the smell's baseline implementation and the author's original results.
\emph{Hermans Feature Envy} illustrates the evaluation result of the original smell detection technique as published by Hermans~\etal \cite{HermansPD12}.
\emph{Baseline Feature Envy} is the result of our baseline interpretation of the smell in \fritz.
\emph{Group Feature Envy} depicts the result of the improved smell detection process as described in Section~\ref{sec:improvedFeatureEnvy}.


In terms of numbers, \emph{Baseline Feature Envy} reports \numprint{3058758} inter-worksheet connections (\numprint{505,4} connections per worksheet).
In comparison, \emph{Group Feature Envy} counts \numprint{102694} inter-worksheet group-connections (\numprint{16,9} connections per worksheet). 
When applying the threshold of 7 for a high risk smell detection, the baseline approach detects \numprint{737} worksheets as smelly, whereas the group-based approach only results in \numprint{533} smell detections.
\change{
	Consultation of the smell's original authors revealed that detailed numbers for \emph{Hermans Feature Envy} are no longer available. 
	The presented numbers are thus extracted from the result plots given in \cite{HermansPD12}.
}
	
When comparing significant features of the graph, \emph{Hermans Feature Envy} detects noticeably more worksheets with a significant number of inter-worksheet connections than the other approaches. 
Based on Hermans \textit{et~al.}'s evaluation, about 70\,\% of the worksheets fall in this category, whereas both \emph{Baseline Feature Envy} and \emph{Group Feature Envy} start reporting signficant detections at the 85\,\% threshold.  
\emph{Hermans Feature Envy} also features wider plateaus of percentage-areas where worksheets contain the same number of Feature Envy connections, whereas the other approaches do not report similar features.
Starting at the 90\,\% mark, the results of all approaches follow a similar trend.
However, the maximal values of the approaches are noticeably different:
Our implementations detect a number of worksheets with more than \numprint{1000}~connections, but Hermans~\textit{et al.} reported no findings of this magnitude.

In terms of detection rates, \emph{Baseline Feature Envy} results in a larger number of individual detections than \emph{Group Feature Envy}. 
However, the group-based approach is based on the number of semantically different connections, instead of the total number of connections. 
It is therefore likely that the group-based approach prevents reporting of false positive smell detections.

\subsection{New Smell Detection Techniques}
\begin{figure}[htbp]
	\centering
	\includegraphics[clip, width=1.0\textwidth, trim=2cm 9.5cm 1.5cm 9cm]{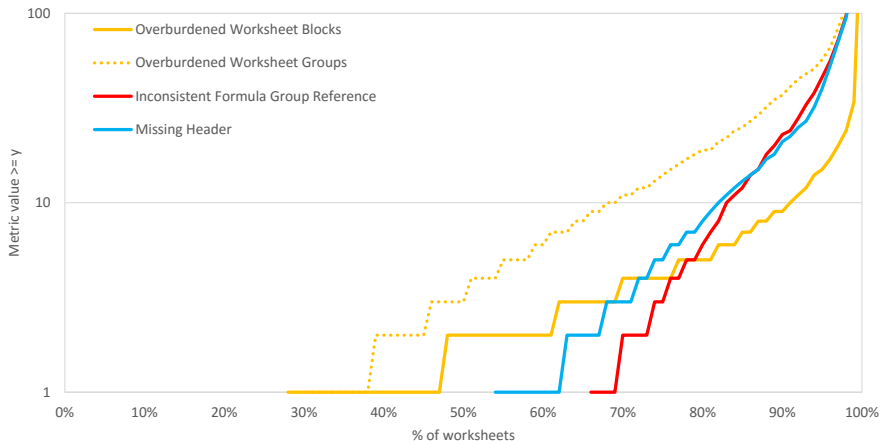}
	\caption{Quartile plot of novel smell metrics}
	\label{fig:figure_eval_novel}
\end{figure}

Figure~\ref{fig:figure_eval_novel} compares the results of our novel smell detection techniques.
\emph{Overburdened Worksheet Blocks} and \emph{Overburdened Worksheet Groups} are smell detection metrics as proposed in Section~\ref{sec:overburdenedWorksheet}, with the former counting the number of calculation blocks, and the latter counting the number of formula groups per worksheet.
\emph{Inconsistent Formula Group Reference} illustrates the results of the corresponding smell as described in Section~\ref{sec:inconsistentFormulaGroupReference}.
\emph{Missing Header} counts the number of missing headers as described in Section~\ref{sec:missingHeader}.

\emph{Overburdened Worksheet Blocks} counts \numprint{24006} blocks (\numprint{3,9} counted per worksheet).
In comparison, \emph{Overburdened Worksheet Groups} counts \numprint{105414} groups for the same worksheets (\numprint{17,4} groups per worksheet).
Moreover, worksheets feature \numprint{102281} \emph{Inconsistent Formula Group Reference}s 
and \numprint{98686} \emph{Missing Header}s. 
The results of all novel smells follow a power law like distribution; 
each individual metric curve has a gentle slope at first and most of its variability on the tail.
This is the usual case for smell metrics, as demonstrated by Hermans~\etal \cite{HermansPD12a}\cite{HermansPD12}.

The values of the metrics for \emph{Overburdened Worksheet Groups} are significantly higher than their block-based counterparts.
This is expected, as blocks aggregate multiple groups.
For smell detection, both versions of the \emph{Overburdened Worksheet} smell require a detection threshold.
Worksheets are indicated as smelly only if this threshold is exceeded by the recorded smell metric.
Following Hermans~\etal's recommendation, we provide threshold values for the 70\,\%, 80\,\%, and 90\,\% marks of the smell metric curves in Table~\ref{table_novelsmells_overburdenedworksheet_thresholds}. 
Worksheets whose metrics surpass these thresholds are regarded as featuring low risk, medium risk, and high risk respectively of being affected by the related smell.

\begin{table*}[thp]
	\centering
	\scriptsize
	\rowcolors{2}{gray!15}{white}
	\begin{tabular}{l|rrr} 
		\toprule
		\multicolumn{1}{c|}{\textbf{Smell Detection Technique}}	& \multicolumn{1}{c}{\textbf{70\,\%}} & \multicolumn{1}{c}{\textbf{80\,\%}} & \multicolumn{1}{c}{\textbf{90\,\%}}\\
		\midrule
		Overburdened Worksheet Blocks	& 4			& 5			& 9\\
		Overburdened Worksheet Groups	& 11		& 19		& 37\\	
		\bottomrule
	\end{tabular}
	\caption{Overburdened Worksheet detection thresholds.}
\label{table_novelsmells_overburdenedworksheet_thresholds}
\end{table*}

\emph{Inconsistent Formula Group Reference} and \emph{Missing Header} smells are reported for each individual instance that is detected within a worksheet, hence no detection thresholds have to be provided. 
\emph{Inconsistent Formula Group Reference} has one or fewer detections for about 70\,\% of the worksheets; \emph{Missing Header} for about 63\,\%. 
Both metrics surpass 10 or fewer detections per worksheet at 80\,\% to 85\,\%.
Hence, about 15\,\% of the worksheets have more than 10 individual detections of the smells.
The upper ends of both metric curves exceed 100 detections per worksheet.
In case of \emph{Inconsistent Formula Group Reference}, these results are probably caused by \spreadsheetfunction{VLOOKUP} and similar spreadsheet functions which refer to entire areas of worksheets instead of individual groups, introducing a large number of inconsistent references.
In case of \emph{Missing Header}, the high metric numbers are likely caused by limitations in the header detection process and the currently applied focus of the header detection method.

\subsection{Manual Investigation}
\change{
To complement the empirical study we described above, we also conducted a manual investigation of detected smells.
For this investigation, we again employed the \emph{homework} category of spreadsheets of Cunha\textit{et~al.}'s evaluation dataset that we previously used to compare the results of the \emph{Pattern Finder} smell in Section~\ref{sec:pattern_finder}.
For each of the basic, improved, and novel smells, we applied the smell detection techniques to all sheets in the collection using the suggested thresholds.
We then tallied how many of each smell instance were detected, how many of the detections were relevant (signalled a genuine issue), and for the basic and improved techniques, how many detections were missing that were successfully indicated by the respective other technique.
}

\change{
Table~\ref{table_manualEvaluation} summarizes the results of this investigation.
In general, significantly less detections were recorded for the improved and novel techniques.
\emph{Pattern Finder} was the most detected smell.
However, many of the detections of the basic smell were spurious, and many of relevant cases that are detected of the improved version were missed by the basic implementation.
The 9 \emph{Missing} cases for the \emph{Improved Pattern Finder} were multiplicities of the same structural issue of one worksheet.
The \emph{Long Calculation Chain} smell was detected in one spreadsheet only (see example below).
Also, we found no detections for both versions of the \emph{Feature Envy} smell, as the spreadsheets in this category do not sufficiently rely on inter-worksheet references.
For the \emph{New Techniques}, almost all detections were relevant and pointed out novel issues.
}

\begin{table*}[thp]
	\centering
	\scriptsize
	\rowcolors{2}{gray!15}{white}
	\begin{tabular}{l|rrr} 
		\toprule
		\multicolumn{1}{c|}{\textbf{Smell Detection Technique}}	& \multicolumn{1}{c}{\textbf{Detected}} & \multicolumn{1}{c}{\textbf{Relevant}} & \multicolumn{1}{c}{\textbf{Missing}}\\
\midrule
\multicolumn{4}{c}{\textit{Basic Techniques}} \vspace{1pt}	\\ 
\midrule
Pattern Finder					& 181		& 20		& 80\\ 
Long Calculation Chain			& 99		& 99		& 0	\\ 
Feature Envy					& 0			& 0			& 0	\\ 
\midrule
\multicolumn{4}{c}{\textit{Improved Techniques}} \vspace{1pt}	\\
\midrule
Pattern Finder					& 9			& 9			& 9	\\ 
Long Calculation Chain			& 0			& 0			& 0	\\ 
Feature Envy					& 0			& 0			& 0	\\ 	
\midrule
\multicolumn{4}{c}{\textit{Novel Techniques}} \vspace{1pt}	\\
\midrule
Overburdened Worksheet 			& 12		& 9			& -\\	
Incons. Ref.					& 12		& 12		& -\\
Missing Header					& 5			& 5			& -\\ 
	\bottomrule
\end{tabular}
	\caption{Manual evaluation of detected smells based on the homework folder of Cunha~\textit{et al.}'s evaluation set.}
\label{table_manualEvaluation}
\end{table*}

\begin{figure}[htp]
	\centering
	\begin{subfigure}[b]{1.0\textwidth}
		\includegraphics[width=\textwidth]{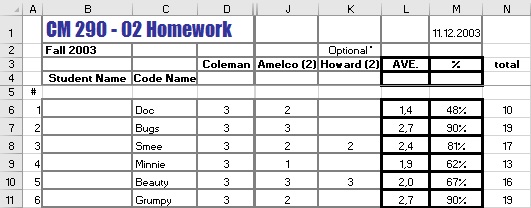}
		\caption{Value view}
		\label{fig:example_homework_1}
	\end{subfigure}
	
	\begin{subfigure}[b]{1.0\textwidth}
		\includegraphics[width=\textwidth]{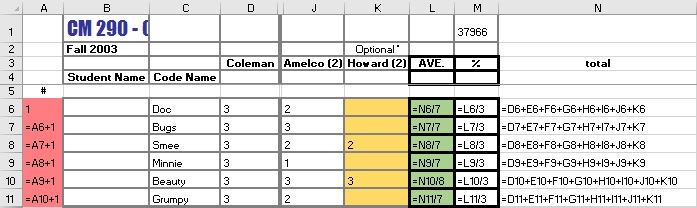}
		\caption{Formula view}
		\label{fig:example_homework_2}
	\end{subfigure}
	\caption{Spreadsheet \texttt{finalGRADES.xls} of Cunha~\textit{et~al.}'s evaluation set.}
	\label{fig:example_homework}
\end{figure}

\change{
Figure~\ref{fig:example_homework} illustrates an excerpt of the \texttt{finalGrades.xls} that was part of the manual investigation. 
This example contains three deficits, that were successfully indicated by spreadsheet smells:
(1)~The student numbers in \texttt{Column A}, after the first entry, are created by successive, self referencing formulas, instead of usually employed continuous numbers.
The issue is highlighted by many \emph{Long Calculation Chain} detections, as each formula after the 7th in the chain is considered smelly.
It is also indicated by one instance of the \emph{Group Pattern Finder} smell for the column.
The basic \emph{Pattern Finder} smell, in contrast, does not detect this issues.
(2)~Many cells in \texttt{Column K} are empty, but are referred to by subsequent calculations, that do not properly take missing values into account.
This issue is, again, highlighted by \emph{Group Pattern Finder} smell, detected for the column, but was not detected by the basic smell version.
(3)~The calculations in \texttt{Column L} use different formulas, dividing the total in \texttt{Column N} bei either 7 or 8. 
The exact number to divide by should likely depend on the optional entry in \texttt{Column K}.
However, closer investigation reveals that this is not properly implemented in the sheet.
The issue is revealed by the \emph{Inconsistent Formula Group Reference} smell, as the percentage calculation in \emph{Column M} refers to a number of smaller \emph{formula groups} in \texttt{Column L} that are created for the different formulas.
Moreover, the issue is also detected by the \emph{Overburdened Worksheet} smell, as many small formula groups are calculated in \texttt{Column L}, which exceeds the threshold for this smell.
}

\subsection{Discussion}

	
\change{
In the empiric evaluation of the \emph{Feature Envy} smell, the baseline interpretation features significantly lower individual detection numbers.
Indeed, it counts \numprint{8003} of \numprint{45010} detections.
The results from the manual investigation suggest that this drop in the detection rate likely excludes a significant portion of the previously detected false positive smell instances.
Further, the \emph{Improved Pattern Finder} detection reveals more smell instances than our chosen baseline (i.e., the non-evaluated version reports \numprint{49667} detections).
This increase of the detection rate is likely attributable to genuine smell detections.
This is also in line with the results from our manual investigation, where the improved technique revealed additional issues that were not detected by the original technique.
}

\change{
When comparing the results of improved \emph{Long Calculation Chain} and \emph{Feature Envy} smells with their baseline implementations, the corresponding metric curves in the empiric evaluation consistently follow a similar trend.
However, the improved versions of the smells indicate overall lower individual metric values. 
Indeed, when applying the suggested threshold values, the improved versions of the smells report a substantially lower number of smell detections than their baseline counterparts (\numprint{4879} instead of \numprint{84031} for Long Calculation Chain, and \numprint{533} instead of \numprint{737} for Feature Envy).
However, genuine detections that would be indicated by the baseline smell are still reported by the improved version.
Our manual investigation also indicates that the improved techniques are successful in limiting the number of superficial detections.
The proposed improvements thus are successful in reducing the number of individual entities that are indicated and have to be checked by a user.
}

Revisiting the first research question (\textbf{RQ1}): \enquote{\rqone}, we conclude that our proposed improvements provide a clear benefit over the original smell detection techniques. 
The improved version of the Pattern Finder detection process decreases the amount of false positives  and \change{reveals new issues}. 
The improved Long Calculation Chain and Feature Envy detection techniques limit duplicate reports, while still including genuine cases.

The metric results of the novel smell detection techniques follow the expected trend set by previous spreadsheet smell evaluations.
The Overburdened Worksheet variants require a threshold to indicate worksheets as smelly.
We provided the cut-off values of four for block-based detection and eleven for group-based detection of the smell, each indicating a low risk for the corresponding worksheet.
Occurrences of inconsistent formula group references and missing headers can directly be reported as smell detections.
\change{As illustrated by our manual investigation, all three smell types could be used alongside the established set of smells and were able to identify novel issues.}

Looking at the second research question (\textbf{RQ2}): \enquote{\rqtwo}, we conclude that the newly introduced smell detection techniques indeed perform similar to traditional spreadsheet ones. Detection values and rates follow the same distribution trend, a power law like distribution which is the usual case for smell metrics (see Hermans~\etal \cite{HermansPD12a}\cite{HermansPD12}).
\change{They are moreover mechanically similar to the existing smells, and successfully point out novel issues, as demonstrated by the manual investigation.}
Consequently, each of the newly introduced smells is apt to be used alongside the currently established smell catalogue.

\change{
Lastly, detection of structure refined smells is dependent upon successful inference of structure information.
However, due to erroneous or unexpected spreadsheet layouts, the proposed structure analysis approach might lead to incomplete or misleading results.
Fortunately, cases of \enquote{improper} spreadsheet structuring usually also cause issues that are detected by structure aware smells.
For example \emph{Inconsistent Formula Group Reference} highlights any structurally unsound modification of a formula group, if the initial group as referenced by another formula group, or if the initial group referred to any other formula group.
A list of examples for such detections is given in previous work \cite{Koch2016}.
If one of these smells identifies and reports the initial structural issue, the user is able to apply adequate fixes and refactorings.
This allows for a bootstrap approach of iterative cycles of structure inference, smell detection, and refactoring, until a sound spreadsheet structure is accomplished.
}

\subsection{Threats to Validity}\label{sec:threatsToValidity}
A threat to the external validity of our evaluation is the representativeness of the EUSES corpus for the overall population of spreadsheets. 
However, the corpus consists of 4490 spreadsheets in 11 categories, providing an adequate variety of samples. Moreover, the corpus has already been extensively used for empirical evaluations, providing further credit as adequate evaluation baseline.

A further threat to the external validity of our results are the preprocessing operations we applied to the corpus before the actual evaluation took place. 
However, this operation only affects the comparison between the results of our baseline implementation and the respective smell's original evaluation. 
Both the evaluation of our baseline implementation as well as the evaluation of improved and new detection techniques are based on the same, preprocessed set of spreadsheets.

A threat to the internal validity of our results might concern the baseline smell detection implementations in the \fritz tool. Abstract smell definitions, provided by each smell's original author, leave room for interpretation for a concrete implementation. Hence, our specific design decisions when implementing the baseline smells might affect the related evaluation results.

Another threat with respect to the internal validity is related to the correctness of our tool implementation, \fritz, providing  spreadsheet file handling, abstraction, structural analysis, smell detection, and automatic evaluation. 
We minimized this risk by manual testing, sanity checking of evaluation results, and comparison of the results with the original evaluation results.
Moreover, the tool is publicly available. 
This allows other researchers to replicate our results.

\section{Related Work}\label{sec:rw}
Hermans \textit{et al.}~\cite{HermansPD12} were among the first to define smells for spreadsheets.
They adapted Fowler's inter-class smells~\cite{Fowler99} from object-oriented software to spreadsheets by treating worksheets as classes:
When two or more worksheets have a strong connection, the spreadsheet is difficult to understand and to maintain;
changes to a worksheet might also have impacts on other worksheets.
In their paper, they redefined well-known smells like \textit{Inappropriate Intimacy}, \textit{Feature Envy}, and \textit{Shotgun Surgery}.
In an ensuing work~\cite{HermansPD12a}, they deal with intra-worksheet smells and  propose smells like \textit{Multiple Operations} (derived from \textit{Long Method} smell), \textit{Multiple References} (from \textit{Long Parameter List}), \textit{Conditional Complexity}, and \textit{Long Calculation Chain} for spreadsheets.

In more recent work, Hermans \textit{et al.} propose refactorings for formula smells~\cite{HermansPD15}.
They indicate the need for refactoring by shading smelly cells and adding comments.
These comments mention the proposed refactoring action (e.g. \enquote{Common subformula can be extracted}).
Hermans and Dig~\cite{sigsoft/HermansD14} provide tool support for refactoring intra-formula smells.
Unfortunately, support for automated refactoring of inter-formula smells is not offered yet, meaning the user has to manually change the spreadsheet when  inter-formula smells like long calculation chains have been detected.

Cunha \textit{et al.}~\cite{CunhaFMMS12,CunhaFRS12} 
focus on input cells and identify, e.g., outliers of numerical values (\textit{Standard Deviation} smell), typos (\textit{String Distance} smell), references to empty cells, mixed use of strings and numerical values in a column (\textit{Pattern Finder} smell), and deviations in data entries (\textit{Quasi-Functional Dependency} smell).
To provide a better overview of existing smells, they aggregate Hermans \textit{et al.}'s smells and their own smells in a catalog and provide a tool named \textit{SmellSheet Detective}\footnote{\label{footnote:smellSheetDetective}download via \url{http://ssaapp.di.uminho.pt/twiki/bin/view/Main/Software}} which implements all of these smells.
Abreu \textit{et al.}~\cite{AbreuCFMPS14} improve the functionality of the \textit{SmellSheet Detective} by combining smell detection with spectrum-based fault localization.
They provide an implementation of their approach in the tool \textit{FaultySheet Detective}\footnotemark[\value{footnote}].

Several approaches detect faults by identifying structures in spreadsheets.
UCheck~\cite{vlc/AbrahamE07} uses header cells as unit information for input and formula cells. 
A unit can be a simple unit like \enquote*{Employee} or a dependent unit like \enquote*{Employee[Anderson]}.
Units can be combined using the \&-operator, e.g., \enquote*{Employee[Anderson]\&Quarter[1]}.
Formula cells inherit their unit from referenced cells.
Since a formula might reference several cells, the resulting units are a combination of the referenced cells' units.
All units must be well-formed.
Violations occur, for example, if two dependent units with the same base unit are combined using an \&-operator, e.g. \enquote*{Employee[Anderson]\&Employee[Bourne]}.
\change{
Im more recent work \cite{CunhaEMS16}, Cunha \textit{et~al.} extended their approach to automatically infer rational schemas from spreadsheets, and to map these schemas to ClassSheets, object-oriented models for spreadsheets that were previously introduced by Engels and Erwig \cite{EngelsE05}.
They also provided and evaluated a catalogue of refactorings for said models, and showed a positive effect on end-users' productivity via an empirical evaluation \cite{CunhaFMMPS16}.
}

Dimension~\cite{chambers:2009} derives dimension information from the headers (e.g., length, time, and speed) and uses the corresponding units (e.g., meter, second, and meter/second) as units for the input cells.
Formula cells inherit their units from the input cells to which they refer to.
Invalid operations (e.g., adding meter and decimeter or meter and meter/second) are reported as errors.

AmCheck~\cite{DouCW14} identifies cell arrays and detects smells based on these arrays.
There are two types of smells that can be detected by AmCheck: the Missing Formula smell and the Inconsistent Formula smell.
The Missing Formula smell occurs in cells which have a constant input value instead of a formula;
the Inconsistent Formula smell occurs in cells whose formulas differ from the formulas of the other cells in the same cell array.

Zhang \textit{et al.}~\cite{Zhang2016} have empirically evaluated UCheck, Dimension, and AmCheck with respect to precision, recall, efficiency, scope, and limitations.
This evaluation shows that AmCheck has the best precision and recall rate and that UCheck and Dimension find different faults compared to AmCheck.

CACheck~\cite{Dou2016} improves over AmCheck by additionally detecting inhomogeneous cell arrays. A row-/column-based cell array is inhomogeneous if it contains a formula cell that references cells in a different column/row.
Furthermore, CACheck removes invalid cell arrays by means of filtering rules.

\textsc{Custodes}~\cite{Cheung:2016} clusters cells by means of strong and weak features.
Strong features are, for example, copy-equivalent formulas and cell dependency patterns;
weak features are the position of the cell within a worksheet, the cell's labels, and the cell's style.
Outlier cells in the individual clusters are identified and classified either as Missing Formula smell or as Dissimilar Formula smell.

Koci \textit{et al.}~\cite{ic3k/KociTRL16} propose a machine learning approach that classifies cells into five categories: headers, attributes (i.e., row headers), meta data (i.e., captions), data, and derived (i.e., content that is derived from the actual data).
Their approach extracts features (e.g., cell type, color, alignment, font type, font style, column- and row index) from the cells and applies different supervised learning techniques (e.g., Random Forest) on them.
In a post-processing step, they detect and repair misclassified cells by means of predefined patterns.

TableCheck~\cite{Dou:2016} detects table clones, i.e., two rectangular blocks of cells which have the same labels.
Table clones are problematic, as they might become inconsistent when a spreadsheet evolves.
TableCheck also reports when detected clones contain inconsistencies like missing and inconsistent formulas.
TableCheck differs from AmCheck, CACheck and \textsc{Custodes}, as it detects inconsistencies between blocks rather than smells within a block.

Amalfitano \textit{et al.}~\cite{AmalfitanoFTSMS14a} propose a reverse engineering process for automatically retrieving data models from spreadsheets.
This process is a top-town approach, meaning that a spreadsheet is decomposed into several worksheets which contain several areas, subareas and sub-subareas.
The decomposition process makes use of the cells' formatting properties for refining the model.
The derived model is visualized as a UML class diagram.
Another interesting work of Amafitano \textit{et al.}~\cite{AmalfitanoSFT16} presents a tool that helps to analyze connections of cells and VBA code.

The range of research publications regarding the overall topic of spreadsheet quality assurance is considerable.
Therefore, we have focused on papers that are closely related to ours:
The last mentioned papers \cite{vlc/AbrahamE07,chambers:2009,DouCW14,Dou2016,Cheung:2016,ic3k/KociTRL16,Dou:2016,AmalfitanoFTSMS14a} deal with the identification of spreadsheet computation structures.
Our structural analysis process builds upon the ideas of UCheck~\cite{vlc/AbrahamE07} and is explained in detail in Section~\ref{sec:structure}.
The first mentioned papers \cite{HermansPD12,HermansPD12a,HermansPD15,CunhaFMMS12,CunhaFRS12} deal with spreadsheet smells.
We discuss how these spreadsheet smells can benefit from the structural analysis in Section~\ref{sec:smellImprovements}.
Since we have limited the discussion of related work to smells and structural analysis, we
refer the interested reader to Jannach \textit{et al.}'s overview paper~\cite{JannachSHW14} for a general overview of quality assurance techniques for spreadsheets.

\section{Conclusion and Future Work}\label{sec:conclusion}

\change{
In this paper, we proposed to compensate present shortcomings of spreadsheet smells by refining smell detection procedures using structural information.
To that end, we first presented an analysis process that infers structural information from spreadsheets.
We then demonstrated smell refinements on the examples of the Pattern Finder, Long Calculation Chain, and Feature Envy smells.
Furthermore, we introduced three new smells that make use of inferred structure information, namely Overburdened Worksheet, Inconsistent Formula Group Reference, and Missing Header.
Empirical evaluation indicated that refined smells indeed have a positive effect on detected smells, and that novel smells are an adequate strategy to indicate further quality deficits.
}

The empirical evaluation shows that the use of structure information improves the performance of the smell detection techniques:
(1)~improved \emph{Pattern Finder} reduces the number of incorrect detections while increasing the number of genuine detections;
(2)~improved \emph{Long Calculation Chain} limits the number of redundant smell reports; and
(3)~improved \emph{Feature Envy refines} the smell's detection focus, reducing the number of detections of permissible cases. 
The evaluation of the new smell detection techniques indicates their applicability along the current smell catalog to detect novel quality issues.

The proposed smell refinements alleviate a major drawback of existing smell detection processes: i.e., that the original approaches often highlight effects instead of causes.
The refined smell detection approach condenses many related smell detections into one;
the user gets a clearer picture of the overall issue and is less focused on an overwhelming number of problems regarding individual cells.
Moreover, the new smells provide additional perspectives for users to assess the overall spreadsheet quality.
In particular, the Overburdened Worksheet smell acts as a good counterbalance to existing formula- and inter-worksheet smells.

\change{
A major drawback of the proposed improved and new smell detection techniques is their reliance on a successful structural analysis process and the associated limited applicability to any given spreadsheet. 
Not all spreadsheets follow the same approach to general spreadsheet structuring, and not all spreadsheets contain applicable formulas, required as cues for the analysis. 
Hence, the structural analysis process in its current form might not always be be applicable and successful, leading to missing or false positive smell detections.
However, detection of structure refined smells also allows a user to tackle this issue by fixing unsound spreadsheet structures using an iterative bootstrap process.
}


\change{
In future work, we want to examine how to best provide adequate representations of structures and related smells to users.
This includes information about the inter-relations between different groups of a spreadsheet via group references, e.g. in form of a graph, which would be a valuable asset for spreadsheet comprehension.
Moreover, smell detection in traditional software development usually is accompanied with a set refactorings, standard transformations of code that remove the indicated issue.
To provide similar actions for spreadsheets, we currently investigate structure-based interactions in spreadsheets that allow us to formulate refactorings for structure-based smells.
Lastly, the presented structure analysis process, as well as the derived smells pose opportunities for future work. For example, by extending the inference of groups to include non-formula related cells, and defining/evaluating further refined and novel smells.
}

\section*{Acknowledgment}
The work described in this paper has been been funded by the Austrian Science Fund (FWF) project {\em DEbugging Of Spreadsheet programs (DEOS)} under contract number I2144 and the Deutsche Forschungsgemeinschaft (DFG) under contract number JA 2095/4-1.


\section*{References}

\bibliographystyle{elsarticle-num}

\end{document}